\documentclass[aps,preprint,showpacs,preprintnumbers,amsmath,amssymb]{revtex4}
\usepackage{amsmath,mathrsfs,amsbsy,color,graphicx,bm,amsthm,amsfonts}
\usepackage{units}
\usepackage{bbm}
\usepackage{times}
\usepackage{dcolumn}
\usepackage{mathrsfs}% Align table columns on decimal point
\usepackage{amsmath,amssymb,epsfig}
\usepackage{amsmath}
\newcommand{\udots}{\mathinner{\mskip1mu\raise1pt\vbox{\kern7pt\hbox{.}}
\mskip2mu\raise4pt\hbox{.}\mskip2mu\raise7pt\hbox{.}\mskip1mu}}
\begin{document}
\title{Does Hawking effect always degrade fidelity of quantum teleportation in Schwarzschild spacetime?}
\author{Shu-Min Wu$^1$\footnote{smwu@lnnu.edu.cn}, Xiao-Wei Fan$^1$\footnote{xwfan0825@163.com (corresponding author)}, Rui-Di Wang$^1$,  Hao-Yu Wu$^1$,   Xiao-Li Huang$^1$\footnote{ huangxiaoli1982@foxmail.com (corresponding author)}, Hao-Sheng Zeng$^2$\footnote{hszeng@hunnu.edu.cn }}
\affiliation{$^1$ Department of Physics, Liaoning Normal University, Dalian 116029, China\\
$^2$ Department of Physics, Hunan Normal University, Changsha 410081, China
}

% \baselineskip=0.65 cm

%\vspace*{0.2cm}
\begin{abstract}
Previous studies have shown that the Hawking effect always destroys quantum correlations and the fidelity of quantum teleportation in the Schwarzschild black hole. Here, we investigate the fidelity of quantum teleportation of Dirac fields between users in Schwarzschild spacetime. We find that, with the increase of the Hawking temperature, the fidelity of quantum teleportation can monotonically increase, monotonically decrease, or non-monotonically increase, depending on the choice of the initial state, which means that the Hawking effect can create net  fidelity of quantum teleportation. This striking result banishes the extended belief that the Hawking effect of the black hole can only destroy the fidelity of quantum teleportation. We also find that quantum steering cannot fully guarantee the fidelity of quantum teleportation  in Schwarzschild spacetime.  This new unexpected source may provide a new idea for the experimental evidence of the Hawking effect.
\end{abstract}

\vspace*{0.5cm}
 \pacs{04.70.Dy, 03.65.Ud,04.62.+v }
\maketitle
\section{Introduction}
Quantum teleportation, first proposed by Bennett $et$ $al.$ \cite{L1}, is one of the most crucial applications for quantum information, which was experimentally proved by Bouwmeester $et$ $al.$ \cite{L2} by using single photons. Quantum teleportation is a basic protocol for transmitting quantum information from one object to another object by shared quantum entanglement, where the spatially separated sender and receiver  can only perform local operations and communicate between themselves via a classical channel \cite{L3}. Quantum teleportation is the foundation of quantum information and an important part of practical quantum technology, which has attracted wide attention \cite{H1,H2,H3,H4,H5,H6,H7,H8,H9,H10}.
In addition, quantum steering, formalized from the viewpoint of quantum information theory \cite{L4,L5}, is a concept first introduced by Schr\"{o}dinger in 1935 \cite{L6,L7}.
Quantum steering refers to the impossibility of describing one party's conditional state by a local hidden state model in the modern view. In other words, quantum steering allows one observer to control a remote subsystem of another observer owned by measuring his subsystem. Thus, quantum steering represents the quantum correlation between quantum entanglement and Bell nonlocality. Unlike quantum entanglement, quantum steering has richer properties in quantum systems, such as two-way steering, one-way steering, and no-way steering, which have been experimentally demonstrated \cite{L8,L9,L10,L11}. Because quantum steering is a crucial resource, the problem of describing the optimal two-qubit state of quantum teleportation under a fixed steering amount is interesting and open \cite{L12}.

From the perspective of Einstein's theory, the gravitational collapse of sufficiently massive stars creates black holes that are fascinating objects in our universe. With the development of astronomy, the existence of black holes has been indirectly or directly confirmed.
For example, (i): the advanced LIGO detector and Virgo detector have detected gravitational waves for the first time in a binary black hole merger system \cite{L13}; (ii): the first image of a supermassive black hole has been taken by the Event Horizon Telescope at the center of the giant elliptical galaxy M87 \cite{L14,L15,L16,L17,L18,L19}; (iii): the Event Horizon Telescope has photographed Sgr A* \cite{L20}. Black hole physics, while making some progress, is still shrouded in mystery, such as the black hole information paradox. As is well-known, the two pillars of modern physics are general relativity and quantum mechanics, of which  the unification remains an open question. In order to solve this contradiction, relativistic quantum information attempts to bridge the gap between general relativity and quantum mechanics. Recently, Pan and his team have used the ``Micius" quantum communication satellite to complete the quantum optical test of the gravitational decoherence effect, indicating that the study of relativistic quantum information has entered the stage of precision experimental verification \cite{L21}.
On the simulation side, there is a general interest in simulating  the Hawking radiation of the black hole and the cosmological particle generation in quantum systems \cite{L22,L23,L24,L25,L26,L27,L28,L29}.
In theory, the Hawking effect of the black hole always has a negative influence on quantum steering, entanglement, discord, coherence, and the fidelity of quantum teleportation of bosonic fields under the case in curved spacetime \cite{L38,L39,L40,L41,L42,L43,L44,L46,L47,L49,L50,L51,L52,L53,L54,L55,L56,L57,QLQ57}. Therefore, one of our motivations is to investigate whether the Hawking effect of the black hole always reduces the fidelity of quantum teleportation of Dirac fields. Another motivation is to discuss whether the Hawking effect has the same effect on quantum steering and the fidelity of quantum teleportation.

In this paper, we investigate quantum teleportation of Dirac fields between users in Schwarzschild spacetime. We assume that Alice and Bob initially share an X-type state and they apply a standard teleportation scheme (STS) to send the unknown state from Alice to Bob. Here, the sender Alice stays stationary at an asymptotically flat region, while the receiver Bob hovers near the event horizon of the black hole. Pan and Jing have found that the fidelity of quantum teleportation of bosonic fields decreases with the increase of the Hawking temperature \cite{L39}. However, we find that the Hawking effect of the black hole has both positive and negative effects on the fidelity of quantum teleportation of Dirac fields; this means that the Hawking effect can not only reduce the fidelity of quantum teleportation but also increase the fidelity in Schwarzschild spacetime. We also find that the influence of the Hawking effect on quantum teleportation is not the same as that on quantum steering, showing that quantum steering cannot guarantee the fidelity of quantum teleportation in curved spacetime.

The structure of the paper is as follows. In Sec. II, we briefly introduce the fidelity of quantum teleportation for the X-type state. In Sec. III, we describe the quantization of Dirac fields in Schwarzschild spacetime. In Sec. IV,  we study  the influence of the Hawking effect on the fidelity of quantum teleportation and quantum steering in Schwarzschild spacetime. The last section is devoted to the summary.
%------------------------------------------------------------------------------------------------------------------------------------------------------------------------------------------------%
\section{The fidelity of quantum teleportation  for X-type state\label{GSCDGE}}
%--------------------------------------------------------------------------------
In this paper, we consider the universally common X-type state of the bipartite system, and its density matrix can be written as
\begin{eqnarray}\label{S1}
\rho_X= \left(\!\!\begin{array}{cccc}
\rho_{11}&0&0&-\rho_{14}\\
0&\rho_{22}&-\rho_{23}&0\\
0&-\rho_{23}&\rho_{33}&0\\
-\rho_{14}&0&0&\rho_{44}
\end{array}\!\!\right).
\end{eqnarray}
Eq.(\ref{S1}) describes the effective quantum states satisfying the unit trace and positive conditions:$\rho_{11}+\rho_{22}+\rho_{33}+\rho_{44}=1$, $\rho_{22}\rho_{33}\geqslant|\rho_{23}|^2$, and $\rho_{11}\rho_{44}\geqslant|\rho_{14}|^2$. If $\rho_{22}\rho_{33}<|\rho_{14}|^2$ or $\rho_{11}\rho_{44}<|\rho_{23}|^2$, the X-type state is entangled.

In general teleportation scheme, we assume that Alice and Bob  initially share an X-type state $\rho_{X}$  in an asymptotically flat region. The unknown pure state that can be teleported from Alice to Bob is represented by $|\phi\rangle$.
Alice and Bob can use some trace-preserving and local quantum operations and classical communication (LOCC) operations for their respective systems.
After these operations, the final state of Bob takes the form
$$\rho_{B}={\rm Tr}_{A,C}[M(|\phi\rangle\langle\phi|\otimes\rho_{X})],$$
where $M$ denotes the trace-preserving LOCC operation.

Note that the dimension of Hilbert space $H_{A}\otimes H_{B}=C^d\otimes C^d$ is $d$.
Therefore, the fidelity of quantum teleportation that is considered as  a measure of the quality of quantum teleportation reads \cite{L58}
\begin{eqnarray}\label{S2}
F=\langle\phi|\rho_{B}|\phi\rangle=\frac{fd+1}{d+1},
\end{eqnarray}
where $f$ is the fully entangled fraction. The fidelity of quantum teleportation achievable can be entirely decided by the fully entangled fraction of the bipartite state in the STS, which is written as \cite{L59}
\begin{eqnarray}\label{S3}
f(\rho)=\max_{\varphi}\langle\varphi|\rho|\varphi\rangle,
\end{eqnarray}
where $|\varphi\rangle$ covers all maximally entangled states.
STS requires the X-type state to meet the condition  $f>1/d$  for providing better fidelity than classical communication, i.e., the quantum region.
In this paper, we only focus on this state in this region.

If the elements of the density matrix in Eq.(\ref{S1}) satisfy the conditions $\rho_{22}+\rho_{33}\geqslant\frac{1}{2}$ and $\rho_{23}\geqslant\frac{1}{2}(1-\rho_{22}-\rho_{33})$, the fully entangled fraction can be expressed as  \cite{L60}
\begin{eqnarray}\label{S4}
f(\rho_{X})=\frac{1}{2}(\rho_{22}+\rho_{33}+2\rho_{23})\geqslant\frac{1}{2}.
\end{eqnarray}
Because we only pay attention to the quantum region with $f>1/d=1/2$, the above conditions are supposed for the X-type state in the following.

%------------------------------------------------------------------------------------------------------------------------------------------------------------------------------------------------%
\section{Quantization of Dirac fields in Schwarzschild spcetime }

Firstly, we briefly review the vacuum structure of Dirac particles in Schwarzschild spacetime. The metric of Schwarzschild spacetime can be given as \cite{L40}
\begin{eqnarray}\label{S5}
ds^2&=&-(1-\frac{2M}{r}) dt^2+(1-\frac{2M}{r})^{-1} dr^2\nonumber\\&&+r^2(d\theta^2
+\sin^2\theta d\varphi^2),
\end{eqnarray}
where $M$ and $r$ are the mass and radius of the black hole, respectively.
We take $c, G, \hbar$ and $k$ as unity for simplicity in this paper.
The Dirac equation \cite{L62} $[\gamma^a e_a{}^\mu(\partial_\mu+\Gamma_\mu)]\Phi=0$ in Schwarzschild spacetime  can be expressed as follows
\begin{eqnarray}\label{S6}
&&-\frac{\gamma_0}{\sqrt{1-\frac{2M}{r}}}\frac{\partial \Phi}{\partial t}+\gamma_1\sqrt{1-\frac{2M}{r}}\bigg[\frac{\partial}{\partial r}+\frac{1}{r}+\frac{M}{2r(r-2M)} \bigg]\Phi \nonumber\\
&&+\frac{\gamma_2}{r}(\frac{\partial}{\partial \theta}+\frac{\cot \theta}{2})\Phi+\frac{\gamma_3}{r\sin\theta}\frac{\partial\Phi}{\partial\varphi}=0,
\end{eqnarray}
where $\gamma_i$ ($i=0,1,2,3$) represent the Dirac matrices \cite{L63,L64}.
Having solved the Dirac equation near the event horizon of the black holes, we gain positive frequency outgoing solutions outside and inside regions of the event horizon as
\begin{eqnarray}\label{S7}
\Phi^+_{{\bold k},{\rm out}}\sim \phi(r) e^{-i\omega u},
\end{eqnarray}
\begin{eqnarray}\label{S8}
\Phi^+_{{\bold k},{\rm in}}\sim \phi(r) e^{i\omega u},
\end{eqnarray}
where $\phi(r)$ represents four-component Dirac spinor, the retarded coordinate $u=t-r_{*}$ with the tortoise coordinate $r_{*}=r+2M\ln\frac{r-2M}{2M}$ \cite{L63,L64}.
Here, $\omega$ and $\bold k$ represent the frequency and wave vector, respectively, which fulfill $\omega=|\mathbf{k}|$ for the massless Dirac field.
Using Eqs.(\ref{S7}) and (\ref{S8}), the Dirac field $\Phi$ can be expanded as
\begin{eqnarray}\label{S9}
\Phi&=&\int d\bold k[\hat{a}^{\rm out}_{\bold k}\Phi^{+}_{\rm out, \bold k}+\hat{b}^{\rm out\dag}_{\bold -k}\Phi^{-}_{\rm out, \bold k}+\hat{a}^{\rm in}_{\bold k}\Phi^{+}_{\rm in, \bold k}+\hat{b}^{\rm in\dag}_{\bold -k}\Phi^{-}_{\rm in, \bold k}],
\end{eqnarray}
where $a^{\rm out}_{\bold k}$ and $b^{\rm out\dag}_{\bold k}$ are the fermionic annihilation and antifermionic creation operators which correspond to the state in the exterior regions of the event horizon, respectively, and $a^{\rm in}_{\bold k}$ and $b^{\rm in\dag}_{\bold k}$ are the fermionic annihilation and antifermionic creation operators which correspond to the state in the interior regions of the event horizon, respectively \cite{L42,L50}.

According to the suggestion of Domour-Ruffini, we can use Kruskal modes to make analytic continuations for Eqs.(\ref{S7}) and (\ref{S8}) \cite{HS1}.
However,  the Kruskal observer can be free to create excitations in any accessible mode. Therefore, the single-frequency Kruskal mode cannot be mapped to a group of single-frequency Schwarzschild modes \cite{HL1}.
To avoid this incongruity, we can adopt the Unruh mode \cite{HL1,HL2,HL3,HL4,HL5}, which provides an intermediate bridge between the Kruskal and Schwarzschild modes. The Unruh operators have the simple Bogoliubov transformations with Schwarzschild modes, which take the forms as
\begin{eqnarray}\label{S10}
\tilde{c}_{\bold k,R}&=&\frac{1}{\sqrt{e^{-\frac{\omega}{T}}+1}}\hat{a}^{\rm out}_{\bold k}-\frac{1}{\sqrt{e^{\frac{\omega}{T}}+1}}\hat{b}^{\rm in\dag}_{\bold {-k}},\nonumber\\
\tilde{c}_{\bold k,L}&=&\frac{1}{\sqrt{e^{-\frac{\omega}{T}}+1}}\hat{a}^{\rm in}_{\bold k}-\frac{1}{\sqrt{e^{\frac{\omega}{T}}+1}}\hat{b}^{\rm out\dag}_{\bold {-k}},\nonumber\\
\tilde{c}^{\dag}_{\bold k,R}&=&\frac{1}{\sqrt{e^{-\frac{\omega}{T}}+1}}\hat{a}^{\rm out\dag}_{\bold k}-\frac{1}{\sqrt{e^{\frac{\omega}{T}}+1}}\hat{b}^{\rm in}_{\bold {-k}},\nonumber\\
\tilde{c}^{\dag}_{\bold k,L}&=&\frac{1}{\sqrt{e^{-\frac{\omega}{T}}+1}}\hat{a}^{\rm in\dag}_{\bold k}-\frac{1}{\sqrt{e^{\frac{\omega}{T}}+1}}\hat{b}^{\rm out}_{\bold {-k}}.
\end{eqnarray}
Here, the subscripts $R$ and $L$ represent the "right" and "left" modes, respectively. Using the operator ordering $\hat{a}^{\rm out}_{\bold k}\hat{b}^{\rm in}_{\bold {-k}}\hat{b}^{\rm out}_{\bold {-k}}\hat{a}^{\rm in}_{\bold k}$, the Unruh vacuum is given by
\begin{eqnarray}\label{S11}
|0\rangle_{\rm U}&=&\frac{1}{e^{-\frac{\omega}{T}}+1}|0000\rangle-\frac{1}{\sqrt{e^{\frac{\omega}{T}}+e^{-\frac{\omega}{T}}+2}}|0101\rangle\nonumber\\
&&+\frac{1}{\sqrt{e^{\frac{\omega}{T}}+e^{-\frac{\omega}{T}}+2}}|1010\rangle-\frac{1}{e^{\frac{\omega}{T}}+1}|1111\rangle,
\end{eqnarray}
where $T=\frac{1}{8\pi M}$ is the Hawking temperature \cite{HL6}, and $|mm'n'n\rangle=|m_{\bold k}\rangle^{+}_{\rm out}|m'_{\bold {-k}}\rangle^{-}_{\rm in}|n'_{\bold {-k}}\rangle^{-}_{\rm out}|n_{\bold {k}}\rangle^{+}_{\rm in}$. Here, $\{|n_{\bold {k}}\rangle^{+}_{\rm out}\}$ and $\{|n_{\bold {-k}}\rangle^{-}_{\rm in}\}$ are the orthonormal bases for the outside and inside regions of the Schwarzschild black hole, respectively. The  superscript $\{+,-\}$ represents the fermion and antifermion.
For the Schwarzschild observer hovering outside the event horizon, the Hawking radiation spectrum from the perspective of an outside observer can be written as $N_F==\frac{1}{e^{\frac{\omega}{T}}+1}$ \cite{L64}.
We can see that the Unruh vacuum observed by the Schwarzschild observer would be detected as a number of the generated fermions $N_F$ corresponding to a thermal Fermion-Dirac statistics of fermions. This is known as the Hawking radiation.
Each fermionic mode has only the first excited state due to the Pauli exclusion principle. The Unruh excited state of the fermionic mode can be expanded as
\begin{eqnarray}\label{S13}
|1\rangle_{\rm U}&=&[q_{R}(\tilde{c}^{\dag}_{\bold k,R}\bigotimes I_{L})+q_{L}(I_{R}\bigotimes\tilde{c}^{\dag}_{\bold k,L})]|0\rangle_{\rm U}\nonumber\\
&=&q_{R}[\frac{1}{\sqrt{e^{-\frac{\omega}{T}}+1}}|1000\rangle-\frac{1}{\sqrt{e^{\frac{\omega}{T}}+1}}|1101\rangle]\nonumber\\
&&+q_{L}[\frac{1}{\sqrt{e^{-\frac{\omega}{T}}+1}}|0001\rangle+\frac{1}{\sqrt{e^{\frac{\omega}{T}}+1}}|1011\rangle],
\end{eqnarray}
with $|q_{R}|^2+|q_{L}|^2=1$.

The operator $\tilde{c}^{\dag}_{\bold k,R}$ in Eq.(\ref{S10}) represents the creation of an antifermion in the interior vacuum and a fermion in the exterior vacuum of the black hole, respectively. Similarly, the operator $\tilde{c}^{\dag}_{\bold k,L}$ in Eq.(\ref{S10}) means that a fermion and an antifermion are created inside and outside the event horizon of the black hole, respectively. Hawking radiation is generated by quantum fluctuations near the event horizon that spontaneously produce pairs of fermion and antifermion.
The fermion and antifermion can radiate toward the inside and outside regions randomly from the event horizon with the total probability $|q_{R}|^2+|q_{L}|^2=1$. Therefore, $q_{R}=1$ represents that all the fermion moves to the outside of the event horizon of the black hole, while all the antifermion moves toward the inside of the event horizon of the black hole. This means that only fermion can be detected as Hawking radiation. Analogously, $q_{L}=1$ means that only antifermion escapes from the event horizon. Therefore, when only fermion (antifermion)  is detected, the single mode approximation for $q_{R}=1$  ($q_{L}=1$) is a special situation. When we discuss our fidelity of quantum teleportation
beyond the single mode approximation,  we investigate different kinds of Unruh modes with
different values of $q_{R}$.

\section{Hawking effect on fidelity of quantum teleportation and quantum steering in Schwarzschild spacetime}
We assume that Alice and Bob initially share an X-type state for two
Unruh modes at an asymptotically flat region of the Schwarzschild black hole. Then, Alice still stays stationary at an asymptotically flat region, while Bob hovers near the event horizon of the black hole. Bob will detect the thermal Fermi-Dirac particle distribution with his excited detector.
According to Eqs.(\ref{S11}) and (\ref{S13}), we can rewrite Eq.(\ref{S1}). Since Bob cannot access the modes inside the event horizon of the black hole, we trace over the inaccessible modes and obtain the reduced density matrix $\rho^{AB_{out}}$ (for detail please see Appendix A).

We assume that Bob's detector is sensitive only to the
fermionic modes, showing that the antifermionic modes cannot be excited in a single detector when a fermion was detected. Therefore, we should trace out the antifermionic mode $\{|n_{\bold {-k}}\rangle^{-}_{\rm out}\}$  outside
the event horizon of the Schwarzschild black hole
\begin{eqnarray}\label{S26}
\rho^{S}_{X}= \left(\!\!\begin{array}{cccccccc}
\rho^S_{11} & 0 & 0 & -\rho^S_{14} \\
0 & \rho^S_{22} & -\rho^S_{23} &0 \\
0 & -\rho^S_{23} & \rho^S_{33} & 0\\
-\rho^S_{14} & 0 & 0 & \rho^S_{44}
\end{array}\!\!\right),
\end{eqnarray}
where
\begin{eqnarray}\label{S27}
\rho^S_{11}&=&(e^{-\frac{\omega}{T}}+1)^{-1}\rho_{11}+|q_{L}|^2(e^{-\frac{\omega}{T}}+1)^{-1}\rho_{22},\nonumber\\
\rho^S_{22}&=&(e^{\frac{\omega}{T}}+1)^{-1}\rho_{11}+[1-|q_{L}|^2(e^{-\frac{\omega}{T}}+1)^{-1}]\rho_{22},\nonumber\\
\rho^S_{33}&=&(e^{-\frac{\omega}{T}}+1)^{-1}\rho_{33}+|q_{L}|^2(e^{-\frac{\omega}{T}}+1)^{-1}\rho_{44},\nonumber\\
\rho^S_{44}&=&(e^{\frac{\omega}{T}}+1)^{-1}\rho_{33}+[1-|q_{L}|^2(e^{-\frac{\omega}{T}}+1)^{-1}]\rho_{44},\nonumber\\
\rho^S_{14}&=&q_{R}(e^{-\frac{\omega}{T}}+1)^{-\frac{1}{2}}\rho_{14},\nonumber\\
\rho^S_{23}&=&q_{R}(e^{-\frac{\omega}{T}}+1)^{-\frac{1}{2}}\rho_{23}.
\end{eqnarray}

We assume that the state $\rho^{S}_{X}$ satisfies the condition $$(e^{\frac{\omega}{T}}+1)^{-1}\rho_{11}+\rho_{22}+(e^{-\frac{\omega}{T}}+1)^{-1}\rho_{33}\geqslant\frac{1}{2}.$$ Therefore, we obtain
\begin{eqnarray}\label{S27}
f(\rho^{S}_{X})&=&\frac{1}{2}\big\{(e^{\frac{\omega}{T}}+1)^{-1}\rho_{11}+\big[1-|q_{L}|^2(e^{-\frac{\omega}{T}}+1)^{-1}\big]\rho_{22}+(e^{-\frac{\omega}{T}}+1)^{-1}\rho_{33}\nonumber\\
&&+|q_{L}|^2(e^{-\frac{\omega}{T}}+1)^{-1}\rho_{44}+2q_{R}(e^{-\frac{\omega}{T}}+1)^{-\frac{1}{2}}\rho_{23}\big\}.
\end{eqnarray}
The change of $f(\rho^{S}_{X})$ related to the Hawking temperature can be expressed as
\begin{eqnarray}\label{S28}
\Delta_{T}f(\rho^{S}_{X}(T))&\equiv&f(\rho^{S}_{X}(T=T_{0}))-f(\rho^{S}_{X}(T=0))\nonumber\\
&=&\frac{1}{2}\big\{(e^{\frac{\omega}{T}}+1)^{-1}\rho_{11}+q^2_{L}(e^{\frac{\omega}{T}}+1)^{-1}\rho_{22}-(e^{\frac{\omega}{T}}+1)^{-1}\rho_{33}\nonumber\\ &&-q^2_{L}(e^{\frac{\omega}{T}}+1)^{-1}\rho_{44}-2q_{R}\rho_{23}\big[1-(e^{-\frac{\omega}{T}}+1)^{-\frac{1}{2}}\big]\big\}.
\end{eqnarray}
The derivative of $f(\rho^{S}_{X})$ with respect to Hawking temperature $T$ can be written as $\frac{\partial f(\rho^{S}_{X})}{\partial T}$.
We can easily obtain that $\frac{\partial f(\rho^{S}_{X})}{\partial T}|_{T_{0}}>0$ means $\Delta f(\rho^{S}_{X}(T_{0}))>0$, and $\Delta f(\rho^{S}_{X}(T_{0}))<0$ means $\frac{\partial f(\rho^{S}_{X})}{\partial T}|_{T_{0}}<0$.

In order to investigate whether quantum steering can guarantee the fidelity of quantum teleportation, we calculate quantum steering from Alice to Bob $S^{A\rightarrow B}(T)$ and  quantum steering from Bob to Alice $S^{B\rightarrow A}(T)$ (for detail please see
Appendix B). There is no ambiguous map  related to the anticommutation properties of field operators in the quantum teleportation and steering for the same initial X-type state $\rho_{X}$ in Schwarzschild spacetime (for detail please see
Appendix C) \cite{HL7,HL8,HL9,HL10,HL11}. In Fig.\ref{Fig1}-\ref{Fig3}, we plot the fully entangled fraction $f(\rho^{S}_{X})$, quantum steering $S^{A\rightarrow B}(T)$ and $S^{B\rightarrow A}(T)$ between two fermions
as a function of the Hawking temperature $T$ for different $\omega$, $q_{R}$, and initial parameters.

\begin{figure}[htbp]
\centering
\includegraphics[height=1.8in,width=2.0in]{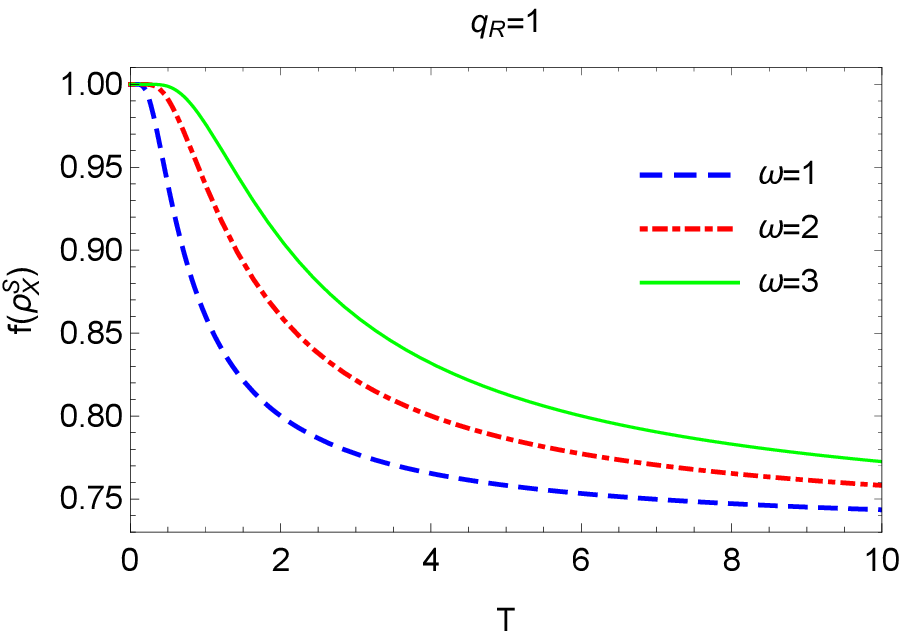}
\includegraphics[height=1.8in,width=2.0in]{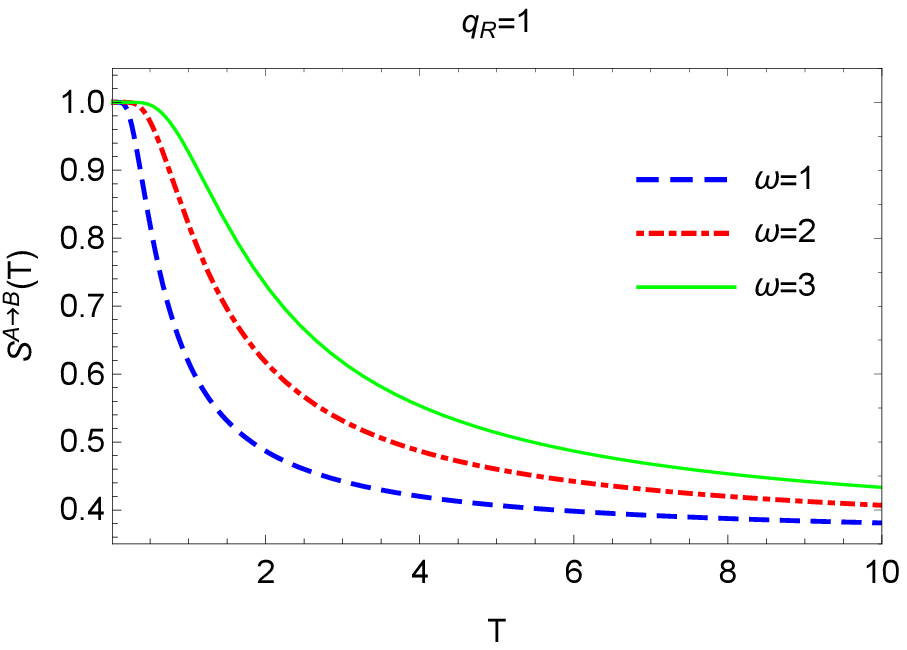}
\includegraphics[height=1.8in,width=2.0in]{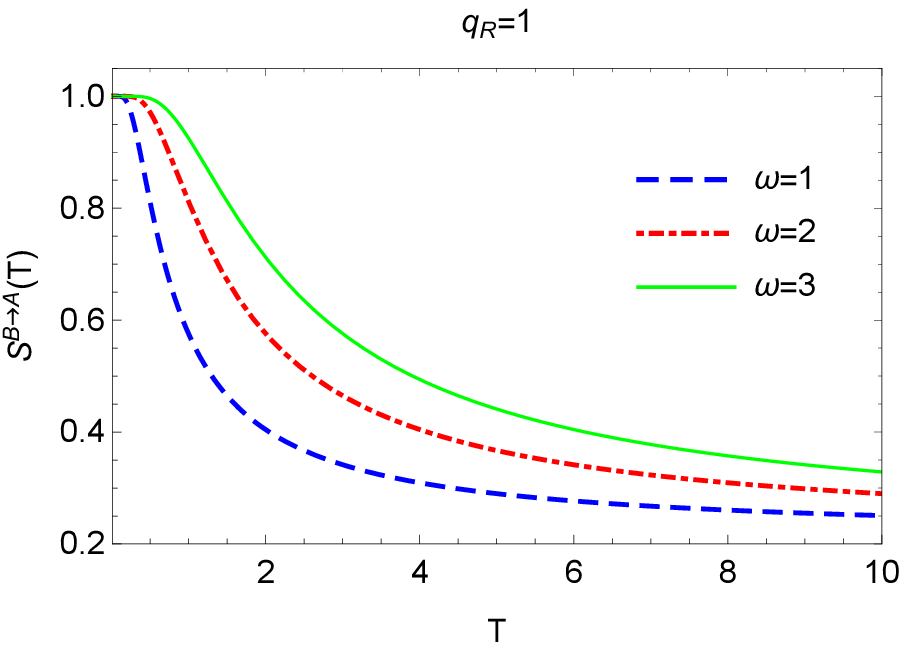}
\includegraphics[height=1.8in,width=2.0in]{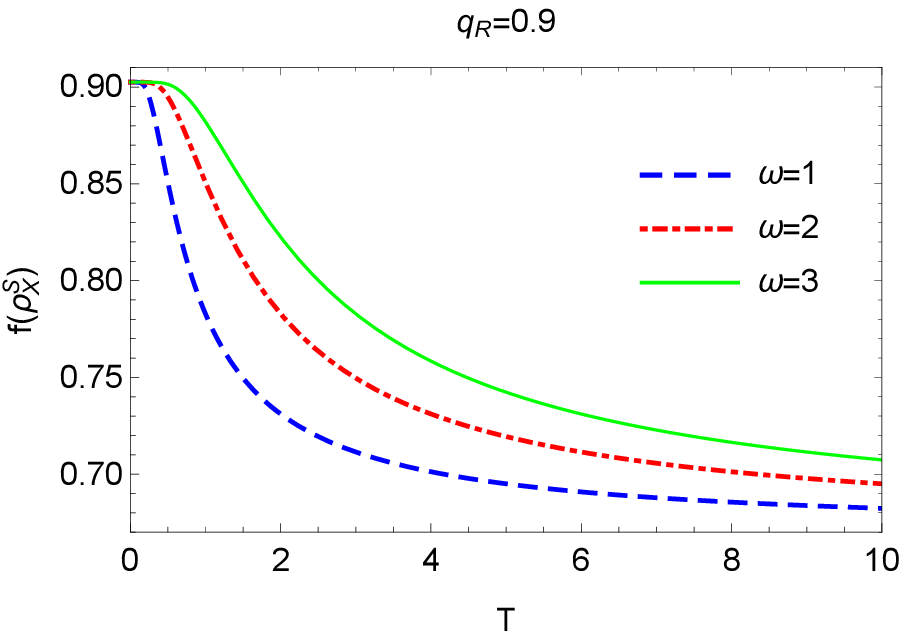}
\includegraphics[height=1.8in,width=2.0in]{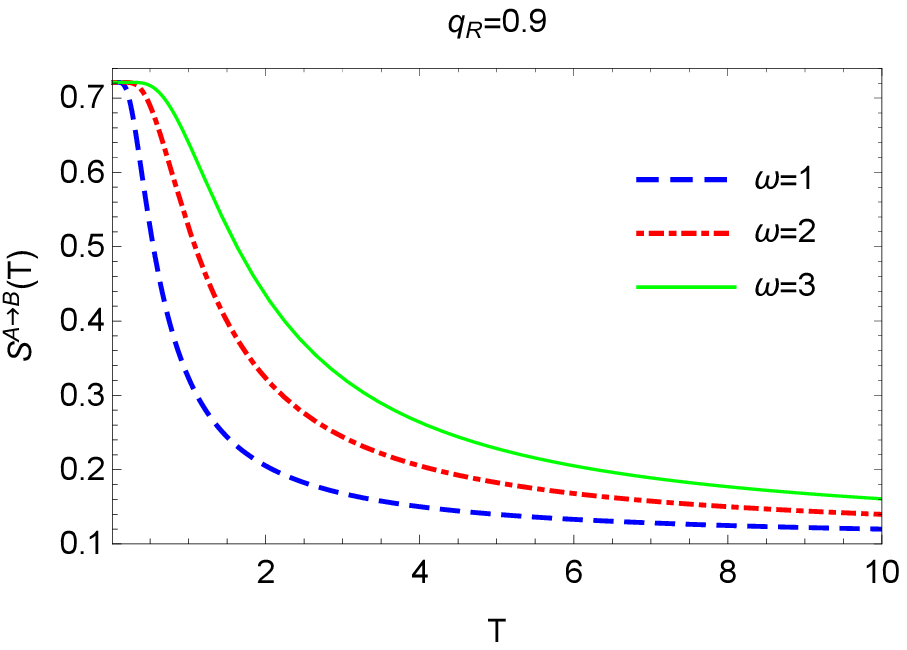}
\includegraphics[height=1.8in,width=2.0in]{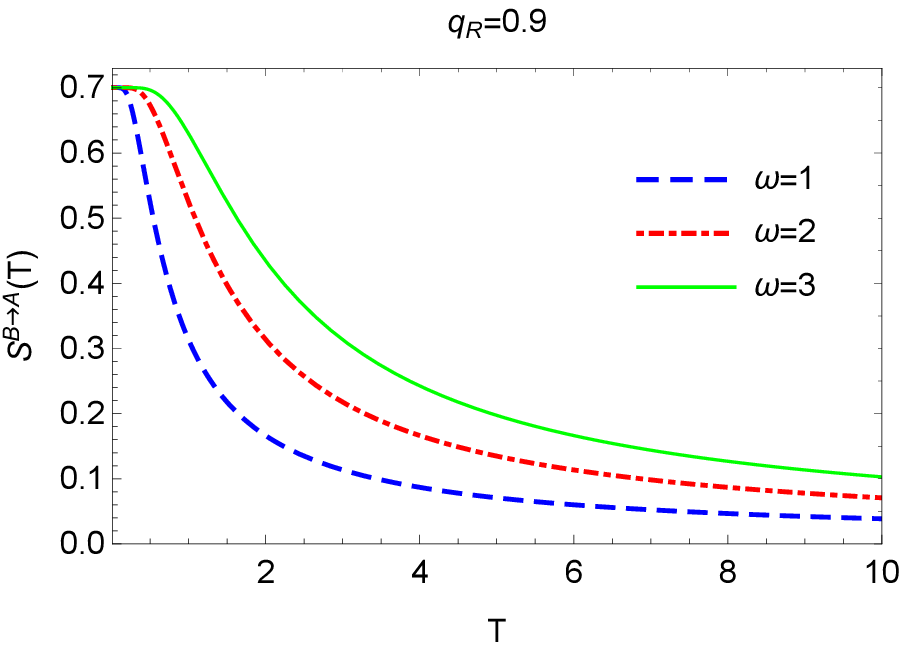}
\includegraphics[height=1.8in,width=2.0in]{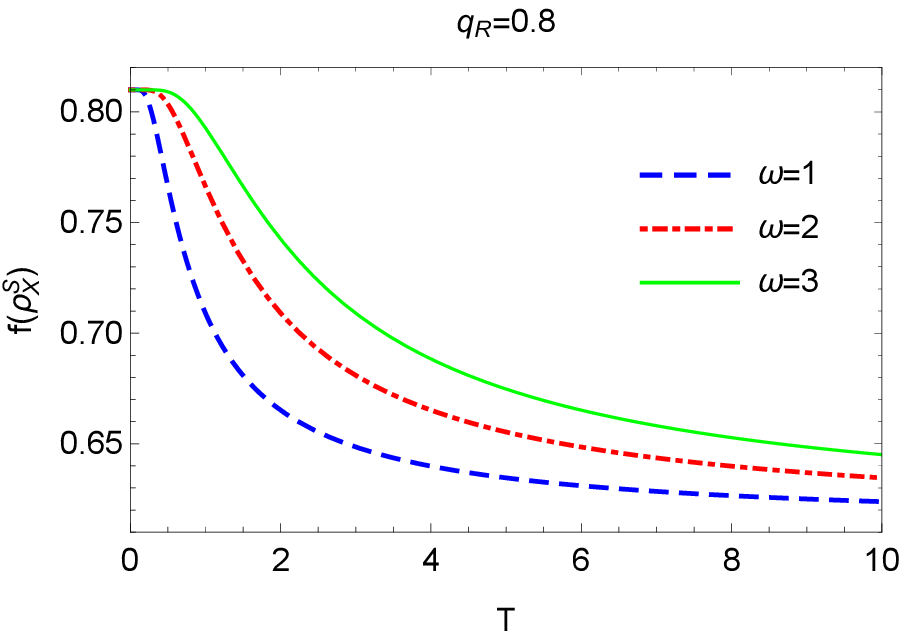}
\includegraphics[height=1.8in,width=2.0in]{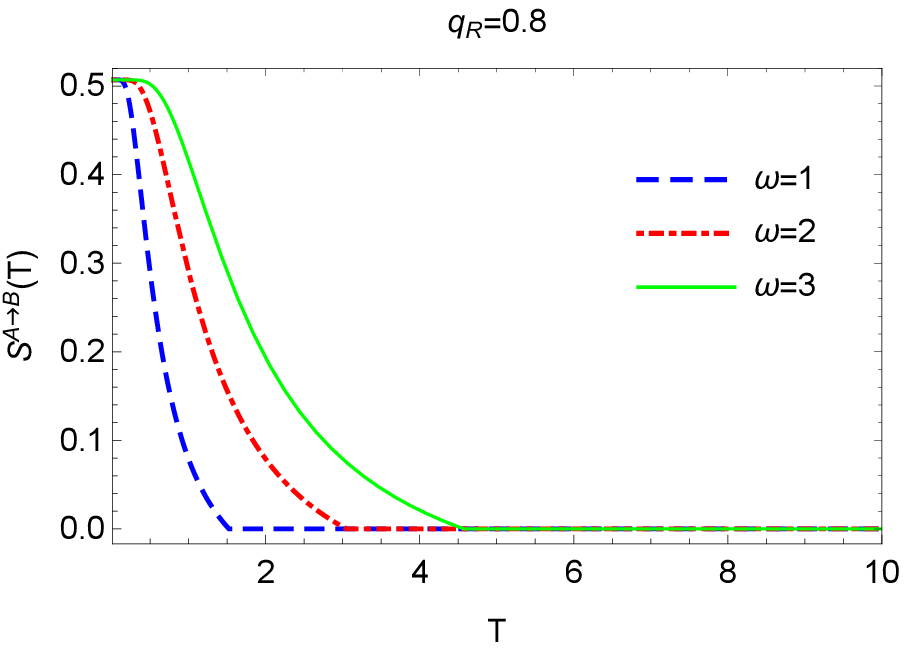}
\includegraphics[height=1.8in,width=2.0in]{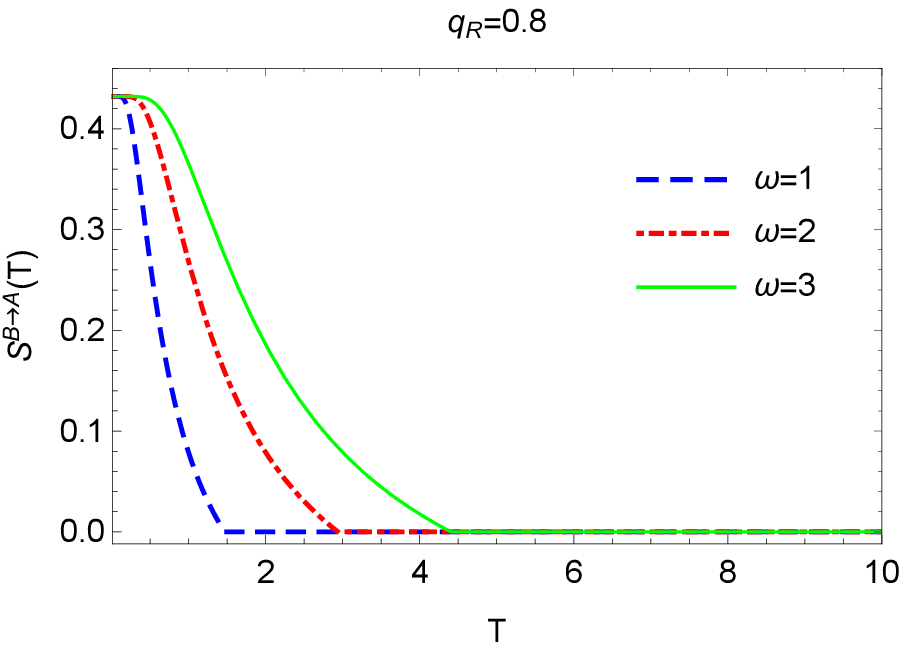}
\caption{ The fully entangled
fraction $f(\rho^{S}_{X})$, quantum steering $S^{A\rightarrow B}(T)$ and $S^{B\rightarrow A}(T)$ as a function of the Hawking temperature $T$ for different $\omega$ and $q_{R}$. The initial parameters are fixed as $\rho_{11}=\rho_{44}=\rho_{14}=0$, and $\rho_{22}=\rho_{33}=\rho_{23}=\frac{1}{2}$.}
\label{Fig1}
\end{figure}
In Fig.\ref{Fig1}, we can see that $f(\rho^{S}_{X})$, $S^{A\rightarrow B}(T)$, and $S^{B\rightarrow A}(T)$ decrease monotonically with the increase of the Hawking temperature $T$. It is worth noting that $f(\rho^{S}_{X})$, $S^{A\rightarrow B}(T)$, and $S^{B\rightarrow A}(T)$ depend on the choice of Unruh modes. For example, quantum steering for  $q_{R}=1$ and  $q_{R}=0.9$ decreases to a fixed value with the Hawking temperature $T$, while quantum steering for $q_{R}=0.8$  suffers from sudden death with $T$. It means that quantum steering cannot fully guarantee the fully entangled fraction in Schwarzschild spacetime. We find that an Unruh mode with $q_{R}=1$  is always optimal to teleport  the unknown pure state to Bob and is optimal for quantum steering between Alice and Bob.  We can also see that the fully entangled fraction and quantum steering are monotonically increasing functions of the frequency $\omega$. The results show that we protect the fully entangled fraction and quantum steering by choosing the high-frequency mode for maximally entangled states
in Schwarzschild spacetime.

\begin{figure}[htbp]
\centering
\includegraphics[height=1.8in,width=2.0in]{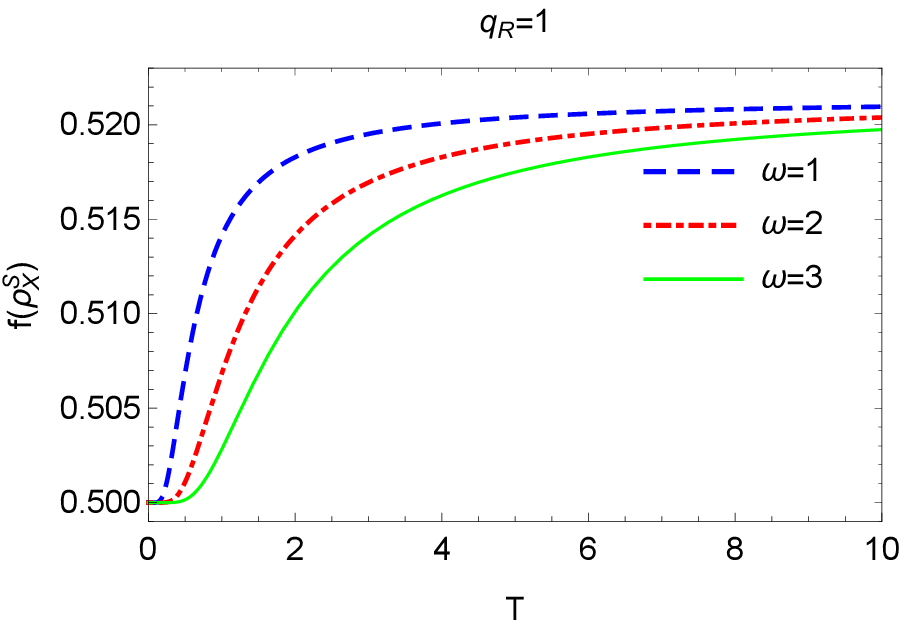}
\includegraphics[height=1.8in,width=2.0in]{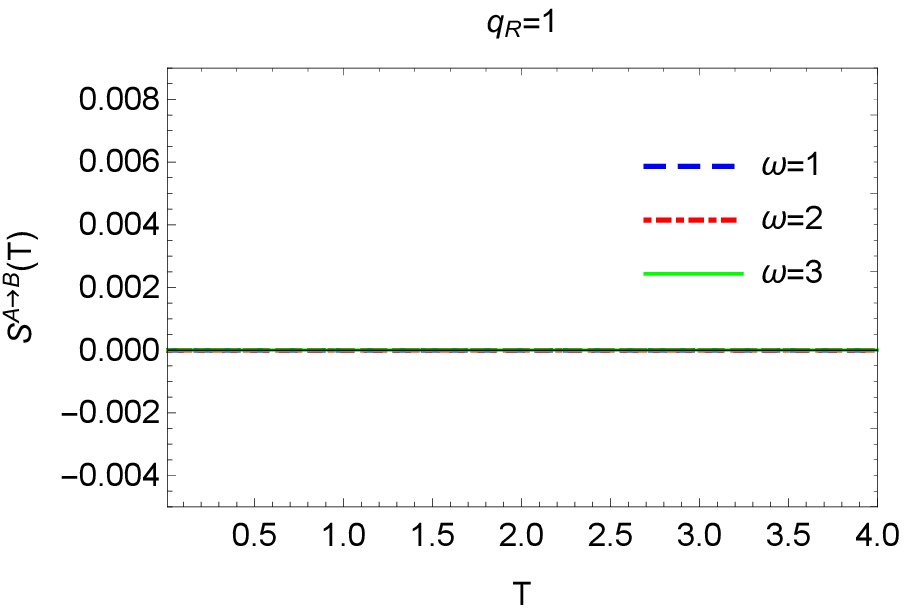}
\includegraphics[height=1.8in,width=2.0in]{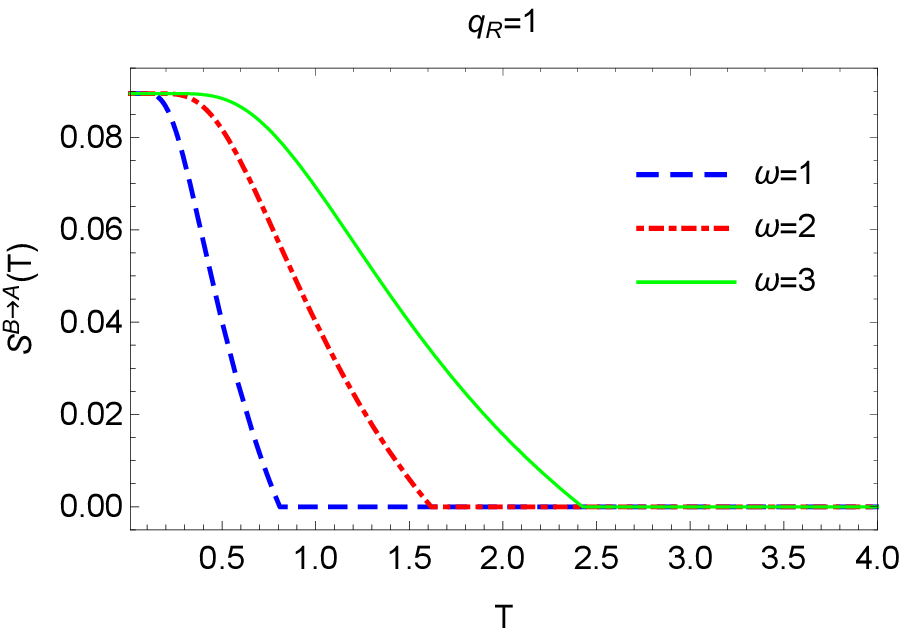}
\includegraphics[height=1.8in,width=2.0in]{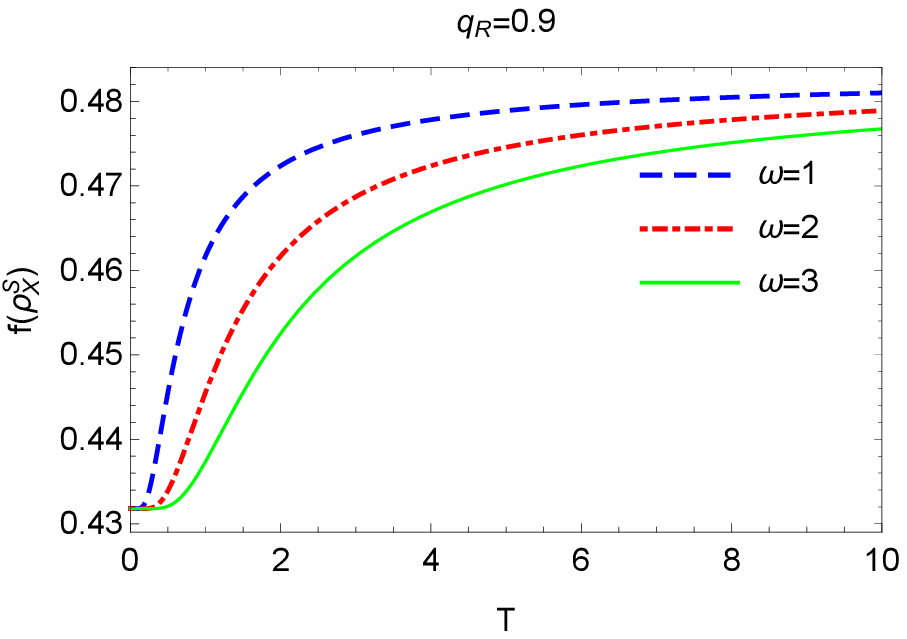}
\includegraphics[height=1.8in,width=2.0in]{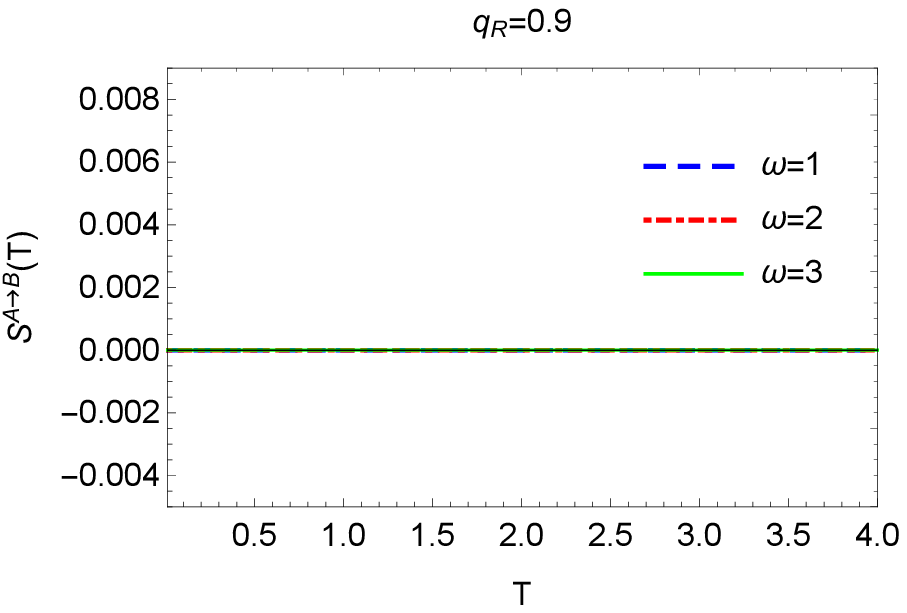}
\includegraphics[height=1.8in,width=2.0in]{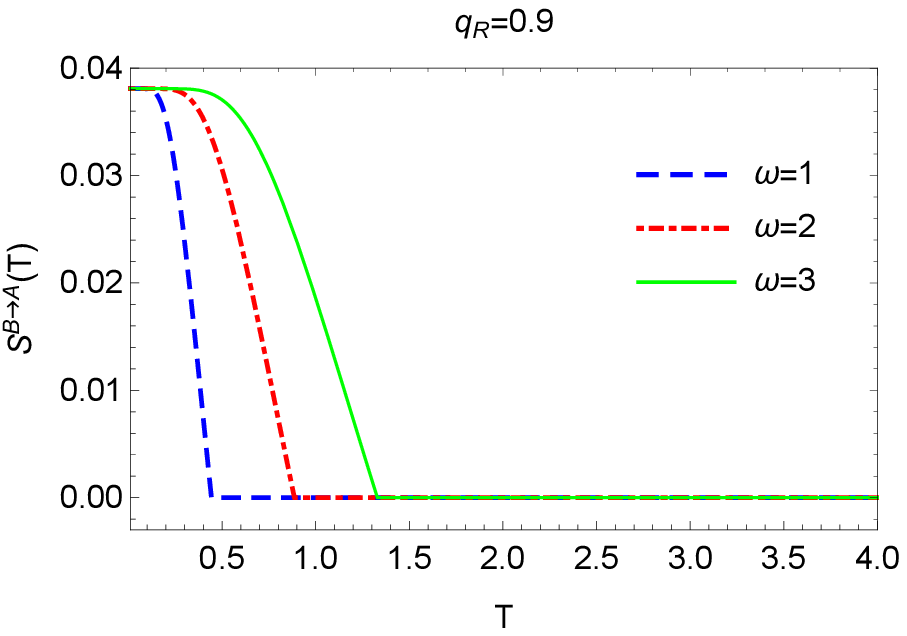}
\includegraphics[height=1.8in,width=2.0in]{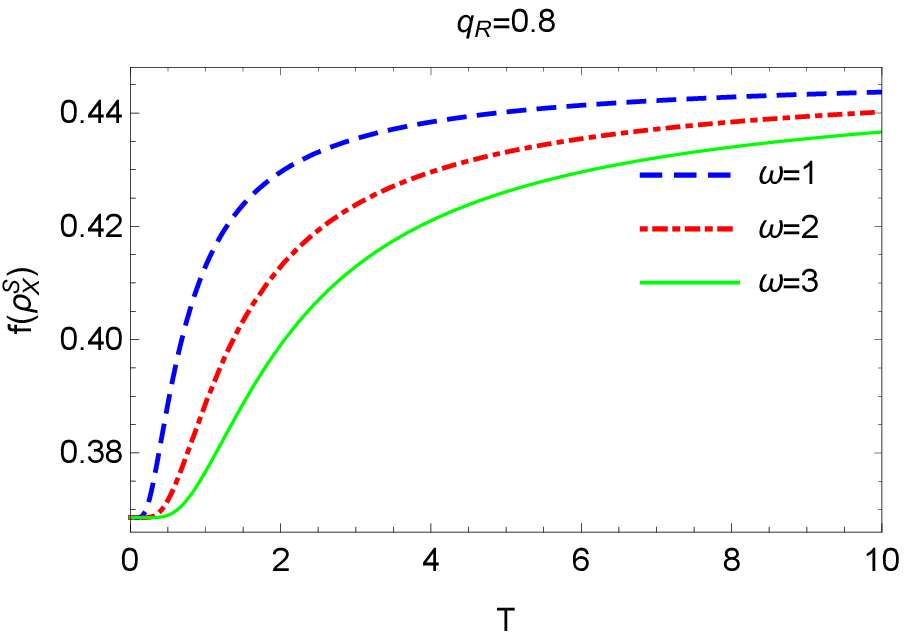}
\includegraphics[height=1.8in,width=2.0in]{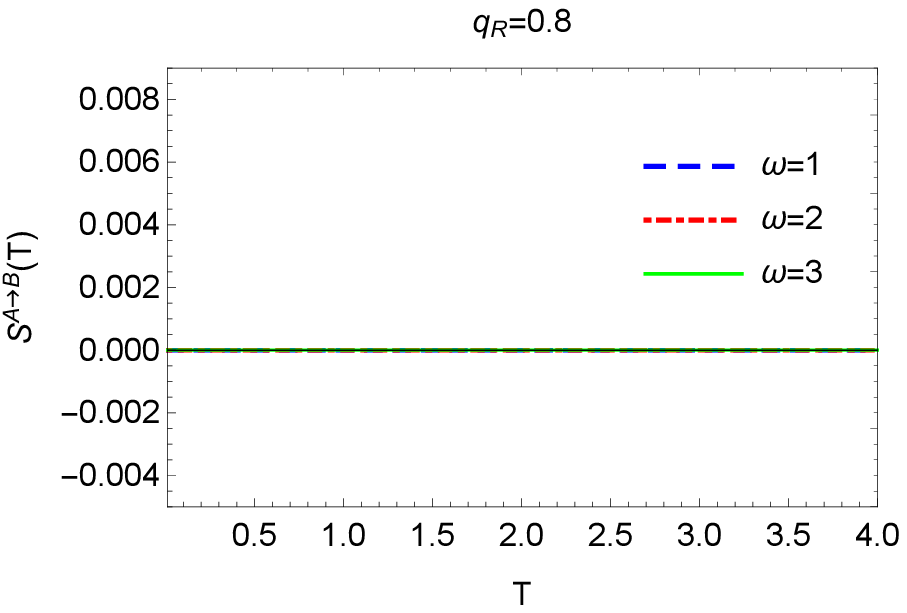}
\includegraphics[height=1.8in,width=2.0in]{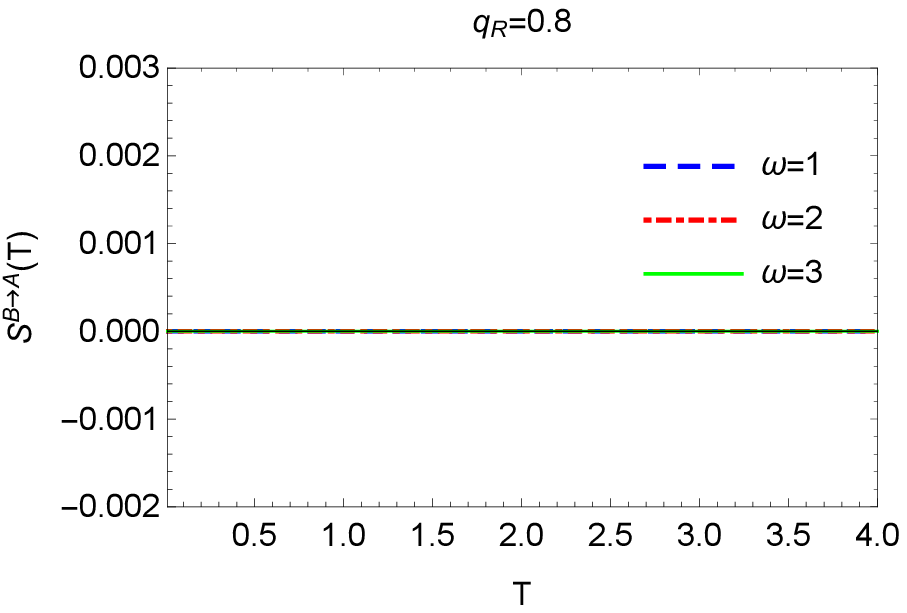}
\caption{ The fully entangled
fraction $f(\rho^{S}_{X})$, quantum steering $S^{A\rightarrow B}(T)$ and $S^{B\rightarrow A}(T)$ as a function of the Hawking temperature $T$ for different  $\omega$ and $q_{R}$. The initial parameters are fixed as $\rho_{11}=\sqrt{2}-1$, $\rho_{22}=\frac{1}{2}$, $\rho_{33}=\frac{3-2\sqrt{2}}{2}$, $\rho_{44}=\rho_{14}=0$, and $\rho_{23}=\frac{\sqrt{2}-1}{2}$.}
\label{Fig2}
\end{figure}
In Fig.\ref{Fig2}, we find that,  with the growth of the Hawking temperature $T$, $f(\rho^{S}_{X})$ increases monotonically, while quantum steering from Alice to Bob $S^{A\rightarrow B}(T)$ is always zero, and quantum steering from Bob to Alice $S^{B\rightarrow A}(T)$ for $q_{R}=1$ and $q_{R}=0.9$ first decreases and then suffers from a ``sudden death". This means that the Hawking effect of the black hole has a positive influence on fully entangled fraction and a negative influence on quantum steering. Therefore, the Hawking effect can create net fidelity of quantum
teleportation, and quantum steering cannot fully guarantee the fully entangled fraction.
However, previous papers have shown that the Hawking effect destroys the fidelity of quantum teleportation and quantum correlation in Schwarzschild spacetime \cite{L38,L39,L40,L41,L42,L43,L44}. Therefore, the Hawking effect of the black hole cannot be simply considered as thermal noise that can only destroy the fidelity of quantum teleportation.
We will use this special type of quantum state to experimentally explore the Hawking effect in the future. We also find that $f(\rho^{S}_{X})$ and $S^{B\rightarrow A}(T)$ increase as $q_{R}$ increases.
In addition, quantum steering from Bob to Alice for $q_{R}=0.8$ is always zero in Schwarzschild spacetime. This again demonstrates that the fully entangled fraction and quantum steering depend on the choice of Unruh modes, and the Unruh mode with $q_{R}=1$ is always optimal for  the fully entangled fraction and quantum steering between Alice and Bob.
Interestingly, increasing the frequency $\omega$ has a negative effect on the fidelity of quantum teleportation and a positive effect on quantum steering for this type of special quantum state. Therefore, we should use low-frequency mode to protect the fidelity of quantum teleportation, while we use quantum steering of high-frequency mode to handle relativistic quantum information tasks. For the first time, we found their different dependence on frequency in Schwarzschild spacetime. These results contribute to our more comprehensive understanding of the Hawking effect of the black hole.

\begin{figure}%[ht]
\centering
\includegraphics[height=1.8in,width=2.0in]{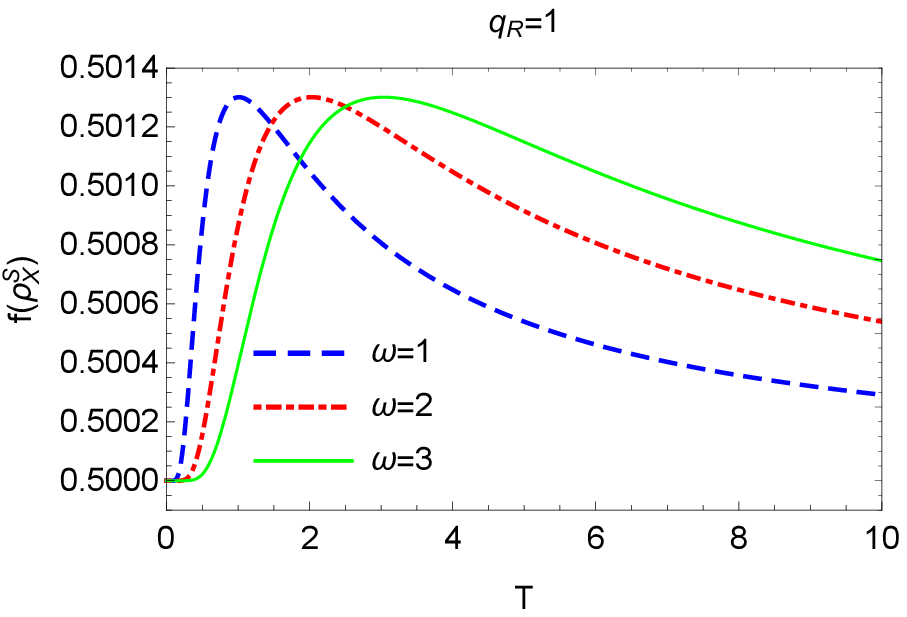}
\includegraphics[height=1.8in,width=2.0in]{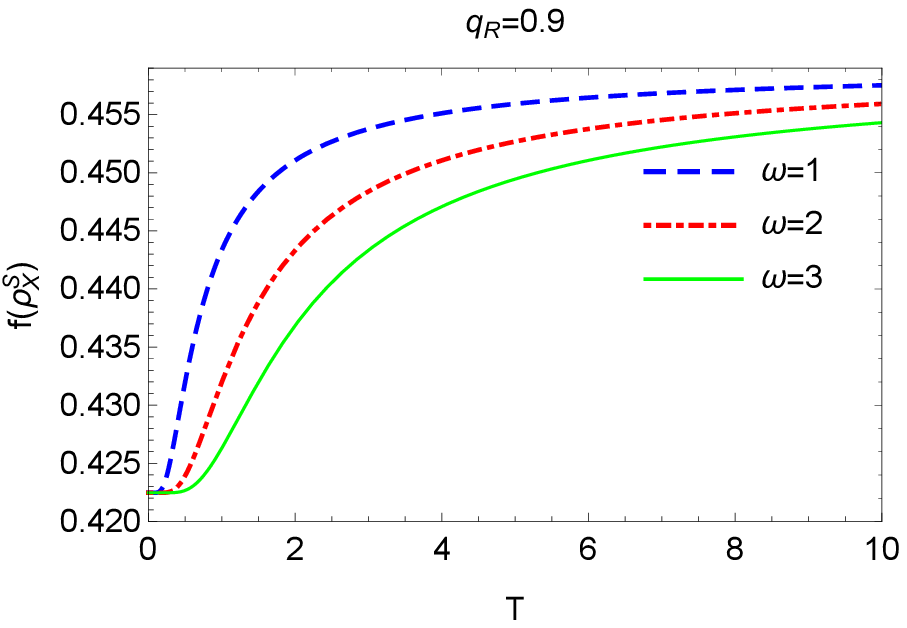}
\includegraphics[height=1.8in,width=2.0in]{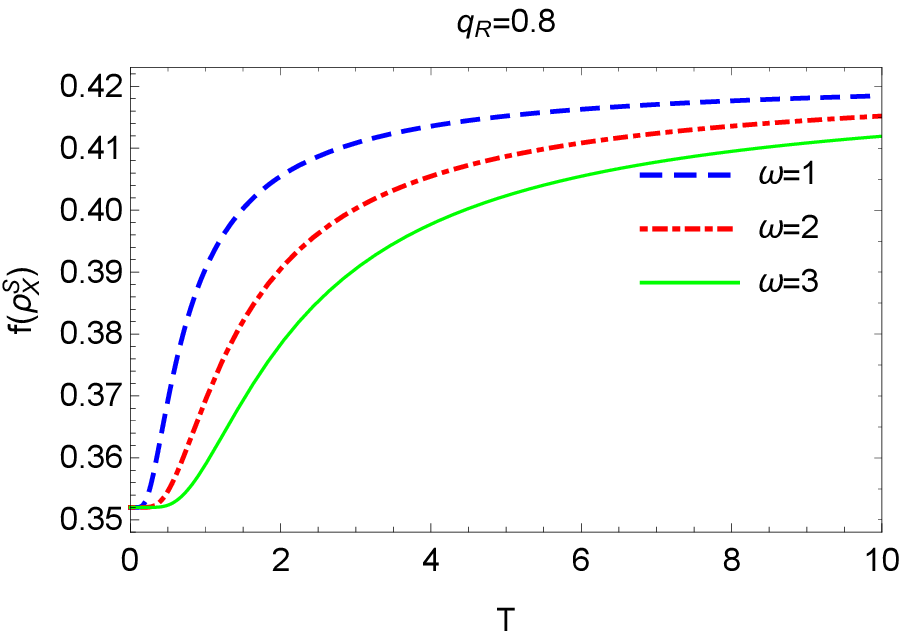}
\caption{The fully entangled fraction $f(\rho^{S}_{X})$ as a function of the Hawking temperature $T$ for different  $\omega$ and $q_{R}$. The initial parameters are fixed as $\rho_{11}=\frac{\sqrt{2}-1}{2}$, $\rho_{22}=\frac{\sqrt{2}}{2}$, $\rho_{33}=\frac{3-2\sqrt{2}}{2}$, $\rho_{44}=\rho_{14}=0$, and $\rho_{23}=\frac{\sqrt{2}-1}{4}$.}
\label{Fig3}
\end{figure}

In Fig.\ref{Fig3}, we find that, for $q_{R}=1$, $f(\rho^{S}_{X})$ first increases from the initial value to the maximum value and then monotonically decreases with the growth of the Hawking temperature $T$. Through the simple calculation, we can obtain $S^{A\rightarrow B}(T)=S^{B\rightarrow A}(T)=0$ in this case. This shows that the Hawking effect has the positive and negative influence on the fully entangled fraction for the
single mode approximation, and quantum steering fully cannot guarantee the fully entangled fraction. We can see that the maximum fidelity of quantum teleportation depends on the Hawking temperature $T$ and frequency $\omega$.
However, for $q_{R}=0.9$ and $q_{R}=0.8$,  $f(\rho^{S}_{X})$ increases monotonically with the increase of the Hawking temperature $T$. For different kinds of Unruh modes, the fully entangled fraction exhibits completely different properties with the Hawking temperature $T$.

%------------------------------------------------------------------------------------------------------------------------------------------------------------------------------------------------%
\section{ Conclutions  \label{GSCDGE}}
%--------------------------------------------------------------------------------
In this paper, we have studied the effect of the Hawking effect on the fidelity of quantum teleportation of Dirac fields between users  beyond the
single-mode approximation in Schwarzschild spacetime. Alice and Bob initially share an X-type state and they apply a standard teleportation scheme (STS) to send the unknown state from Alice to Bob. Here, Alice stays stationary at an asymptotically flat region, while Bob hovers near the event horizon of the black hole. We find that the fidelity of quantum teleportation of Dirac fields can monotonically increase, monotonically decrease, or non-monotonically increase, depending on the choice of the initial state with the increase of the Hawking temperature, meaning that the Hawking effect can enhance and create net fidelity of quantum teleportation.
This makes sharp a contrast with quantum correlation (quantum steering, entanglement, and discord) and  the fidelity of quantum teleportation of bosonic fields, which decrease monotonically with the growth of the Hawking temperature in Schwarzschild spacetime \cite{L38,L39,L40,L41,L42,L43,L44}. The reduction of physically accessible fidelity of quantum teleportation  and steering by the Hawking effect  can be attributed to the increase of physically inaccessible fidelity of quantum teleportation  and steering by the Hawking effect.

In addition, the fidelity of quantum teleportation and quantum steering depend on the choice of Unruh modes.  We showed that the Unruh mode with $q_{R}=1$  is always optimal to teleport  the unknown pure state to Bob  and
$q_{R}=0$ is optimal for quantum teleportation with anti-Bob inside the event horizon of the black hole \cite{HL2}.
For different kinds of Unruh modes, the fidelity of quantum teleportation exhibits completely different properties with the Hawking temperature in curved spacetime (please refer to Fig.\ref{Fig3} for detail). We also find that quantum steering cannot guarantee the fidelity of quantum teleportation in Schwarzschild spacetime. Interestingly, the low-frequency mode may be beneficial for protecting the fidelity of quantum teleportation and harmful to quantum steering.
These surprising results overturn the extended belief that the Hawking effect of the black hole can only destroy the fidelity of quantum teleportation and provides a new and unexpected source for finding experimental evidence of the Hawking effect in curved spacetime.
\appendix
%\begin{widetext}
\onecolumngrid
\section{$\rho^{AB_{out}}$ }
An X-type state is initially shared by Alice and Bob at an asymptotically flat region.
Then, we let Bob hover near the event horizon of the black hole.
According to Eqs.(\ref{S11}) and (\ref{S13}), we can rewrite Eq.(\ref{S1}).
Since  the exterior region and the interior region are causally disconnected,
we should trace over the inaccessible modes and obtain
\begin{eqnarray}\label{S2PO6}
\rho^{AB_{out}}= \left(\!\!\begin{array}{cccccccc}
\rho^{AB_{out}}_{11}& 0 & 0 & -\rho^{AB_{out}}_{14} &0 &\rho^{AB_{out}}_{16} &-\rho^{AB_{out}}_{17} &0\\
0& \rho^{AB_{out}}_{22} & 0 & 0 &\rho^{AB_{out}}_{25} &0 &0 &-\rho^{AB_{out}}_{28}\\
0& 0 & \rho^{AB_{out}}_{33} & 0 &-\rho^{AB_{out}}_{35} &0 &0 &\rho^{AB_{out}}_{38}\\
-\rho^{AB_{out}}_{14}& 0 & 0 & \rho^{AB_{out}}_{44} &0 &-\rho^{AB_{out}}_{46} &\rho^{AB_{out}}_{47} &0\\
0& \rho^{AB_{out}}_{25} & -\rho^{AB_{out}}_{35} & 0 &\rho^{AB_{out}}_{55} &0 &0 &-\rho^{AB_{out}}_{58}\\
\rho^{AB_{out}}_{16}& 0 & 0 & -\rho^{AB_{out}}_{46} &0 &\rho^{AB_{out}}_{66} &0 &0\\
-\rho^{AB_{out}}_{17}& 0 & 0 & \rho^{AB_{out}}_{47} &0 &0 &\rho^{AB_{out}}_{77} &0\\
0& -\rho^{AB_{out}}_{28} & \rho^{AB_{out}}_{38} & 0 &-\rho^{AB_{out}}_{58} &0 &0 &\rho^{AB_{out}}_{88}\\
\end{array}\!\!\right),
\end{eqnarray}
where
\begin{eqnarray}\label{S27}
\rho^{AB_{out}}_{11}&=&(e^{-\frac{\omega}{T}}+1)^{-2}\rho_{11}+|q_{L}|^2(e^{-\frac{\omega}{T}}+1)^{-1}\rho_{22},\nonumber\\
\rho^{AB_{out}}_{22}&=&(e^{-\frac{\omega}{T}}+1)^{-1}(e^{\frac{\omega}{T}}+1)^{-1}\rho_{11},\nonumber\\ \rho^{AB_{out}}_{33}&=&(e^{-\frac{\omega}{T}}+1)^{-1}(e^{\frac{\omega}{T}}+1)^{-1}\rho_{11}+\big[|q_{R}|^2(e^{-\frac{\omega}{T}}+1)^{-1}+|q_{L}|^2(e^{\frac{\omega}{T}}+1)^{-1}\big]\rho_{22},\nonumber\\ \rho^{AB_{out}}_{44}&=&(e^{\frac{\omega}{T}}+1)^{-2}\rho_{11}+|q_{R}|^2(e^{\frac{\omega}{T}}+1)^{-1}\rho_{22},\nonumber\\
\rho^{AB_{out}}_{55}&=&(e^{-\frac{\omega}{T}}+1)^{-2}\rho_{33}+|q_{L}|^2(e^{-\frac{\omega}{T}}+1)^{-1}\rho_{44},\nonumber\\
\rho^{AB_{out}}_{66}&=&(e^{-\frac{\omega}{T}}+1)^{-1}(e^{\frac{\omega}{T}}+1)^{-1}\rho_{33},\nonumber\\
\rho^{AB_{out}}_{77}&=&(e^{-\frac{\omega}{T}}+1)^{-1}(e^{\frac{\omega}{T}}+1)^{-1}\rho_{33}+\big[|q_{R}|^2(e^{-\frac{\omega}{T}}+1)^{-1}+|q_{L}|^2(e^{\frac{\omega}{T}}+1)^{-1}\big]\rho_{44},\nonumber\\
\rho^{AB_{out}}_{88}&=&(e^{\frac{\omega}{T}}+1)^{-2}\rho_{33}+|q_{R}|^2(e^{\frac{\omega}{T}}+1)^{-1}\rho_{44},\nonumber\\
\rho^{AB_{out}}_{14}&=&q_{R}q_{L}(e^{-\frac{\omega}{T}}+1)^{-\frac{1}{2}}(e^{\frac{\omega}{T}}+1)^{-\frac{1}{2}}\rho_{22},\nonumber\\
\rho^{AB_{out}}_{16}&=&q_{L}(e^{-\frac{\omega}{T}}+1)^{-1}(e^{\frac{\omega}{T}}+1)^{-\frac{1}{2}}\rho_{23},\nonumber\\
\rho^{AB_{out}}_{17}&=&q_{R}(e^{-\frac{\omega}{T}}+1)^{-\frac{3}{2}}\rho_{14},\nonumber\\
\rho^{AB_{out}}_{25}&=&q_{L}(e^{-\frac{\omega}{T}}+1)^{-1}(e^{\frac{\omega}{T}}+1)^{-\frac{1}{2}}\rho_{14},\nonumber\\
\rho^{AB_{out}}_{28}&=&q_{R}(e^{-\frac{\omega}{T}}+1)^{-\frac{1}{2}}(e^{\frac{\omega}{T}}+1)^{-1}\rho_{14},\nonumber\\
\rho^{AB_{out}}_{35}&=&q_{R}(e^{-\frac{\omega}{T}}+1)^{-\frac{3}{2}}\rho_{23},\nonumber\\
\rho^{AB_{out}}_{38}&=&q_{L}(e^{\frac{\omega}{T}}+1)^{-\frac{3}{2}}\rho_{23},\nonumber\\
\rho^{AB_{out}}_{46}&=&q_{R}(e^{\frac{\omega}{T}}+1)^{-1}(e^{-\frac{\omega}{T}}+1)^{-\frac{1}{2}}\rho_{23},\nonumber\\
\rho^{AB_{out}}_{47}&=&q_{L}(e^{\frac{\omega}{T}}+1)^{-\frac{3}{2}}\rho_{14},\nonumber\\
\rho^{AB_{out}}_{58}&=&q_{R}q_{L}(e^{-\frac{\omega}{T}}+1)^{-\frac{1}{2}}(e^{\frac{\omega}{T}}+1)^{-\frac{1}{2}}\rho_{44}\nonumber.
\end{eqnarray}

\section{Quantification of quantum steering for X-type state in Schwarzschild spacetime }
As is well-known, the bipartite entanglement can be effectually measured by the concurrence. For the X-state $\rho_{X}$ of Eq.(\ref{S1}), the concurrence is given by \cite{L66}
\begin{eqnarray}\label{A1}
C(\rho_X)=2\max\{|\rho_{14}|-\sqrt{\rho_{22}\rho_{33}}, |\rho_{23}|-\sqrt{\rho_{11}\rho_{44}}\}.
\end{eqnarray}
For any bipartite state $\rho_{AB}$ between Alice and Bob, quantum steering from Bob to Alice can be recognized if the density matrix $\tau_{AB}^1$ that becomes
\begin{eqnarray}\label{A2}
\tau_{AB}^1=\frac{\rho_{AB}}{\sqrt{3}}+\frac{3-\sqrt{3}}{3}(\rho_A\otimes\frac{I}{2}),
\end{eqnarray}
is entangled \cite{L67,L68}, where  $\rho_A$ is Alice's reduced density matrix, $\rho_A={\rm Tr}_B(\rho_{AB})$, and  $I$ is the two-dimension identity matrix of Bob's subsystem. Analogously, quantum steering from Alice to Bob can be proved if the state $\tau_{AB}^2$  that reads
\begin{eqnarray}\label{A3}
\tau_{AB}^2=\frac{\rho_{AB}}{\sqrt{3}}+\frac{3-\sqrt{3}}{3}(\frac{I}{2}\otimes\rho_B),
\end{eqnarray}
is entangled, where $\rho_B={\rm Tr}_A(\rho_{AB})$.
Therefore, the matrix $\tau_{AB}^{1}$ for the X-state $\rho_{X}$ of Eq.(\ref{S1}) can be specifically expressed as
\begin{eqnarray}\label{A4}
\tau_{AB}^{1,x}= \left(\!\!\begin{array}{cccc}
\frac{\sqrt{3}}{3}\rho_{11}+r&0&0&-\frac{\sqrt{3}}{3}\rho_{14}\\
0&\frac{\sqrt{3}}{3}\rho_{22}+r&-\frac{\sqrt{3}}{3}\rho_{23}&0\\
0&-\frac{\sqrt{3}}{3}\rho_{23}&\frac{\sqrt{3}}{3}\rho_{33}+s&0\\
-\frac{\sqrt{3}}{3}\rho_{14}&0&0&\frac{\sqrt{3}}{3}\rho_{44}+s
\end{array}\!\!\right),
\end{eqnarray}
with $r=\frac{(3-\sqrt{3})}{6}(\rho_{11}+\rho_{22})$ and $s=\frac{(3-\sqrt{3})}{6}(\rho_{33}+\rho_{44})$.
Using Eq.(\ref{S5}), the state $\tau_{AB}^{1,x}$ is entangled as long as one of the conditions $|\rho_{14}|^2>\mathcal{F}_a-\mathcal{F}_b$ and $|\rho_{23}|^2>\mathcal{F}_c-\mathcal{F}_b$ is satisfied, where
\begin{eqnarray}
 \nonumber&&\mathcal{F}_a=\frac{2-\sqrt{3}}{2}\rho_{11}\rho_{44}+\frac{2+\sqrt{3}}{2}\rho_{22}\rho_{33}
+\frac{1}{4}(\rho_{11}+\rho_{44})(\rho_{22}+\rho_{33}),\\ \nonumber
&&\mathcal{F}_b=\frac{1}{4}(\rho_{11}-\rho_{44})(\rho_{22}-\rho_{33}),\\ \nonumber
&& \mathcal{F}_c=\frac{2+\sqrt{3}}{2}\rho_{11}\rho_{44}+\frac{2-\sqrt{3}}{2}\rho_{22}\rho_{33}
+\frac{1}{4}(\rho_{11}+\rho_{44})(\rho_{22}+\rho_{33}) \nonumber.
\end{eqnarray}
Thus the steering from Bob to Alice is proved. Similarly, the steering from Alice to Bob can also be proved by one of the inequalities,
\begin{eqnarray}\label{A5}
|\rho_{14}|^2>\mathcal{F}_a+\mathcal{F}_b,
\end{eqnarray}
or
\begin{eqnarray}\label{A6}
|\rho_{23}|^2>\mathcal{F}_c+\mathcal{F}_b.
\end{eqnarray}

Next, we introduced quantities
\begin{eqnarray}\label{A7}
S^{A\rightarrow B}={\rm{max}}\bigg\{0,\frac{8}{\sqrt{3}}(|\rho_{14}|^2-\mathcal{F}_a-\mathcal{F}_b),
\frac{8}{\sqrt{3}}(|\rho_{23}|^2-\mathcal{F}_c-\mathcal{F}_b)\bigg\},
\end{eqnarray}
and
\begin{eqnarray}\label{A8}
S^{B\rightarrow A}={\rm{max}}\bigg\{0,\frac{8}{\sqrt{3}}(|\rho_{14}|^2-\mathcal{F}_a+\mathcal{F}_b),
\frac{8}{\sqrt{3}}(|\rho_{23}|^2-\mathcal{F}_c+\mathcal{F}_b)\bigg\},
\end{eqnarray}
to quantify the steerability from Alice to Bob and  from Bob to Alice, respectively.
Here, the factor $\frac{8}{\sqrt{3}}$ is to ensure that the maximum steering is $1$.

For the X-type state in Eq.(\ref{S26}), using Eqs.(\ref{A7}) and (\ref{A8}), we obtain the quantum steering from Alice to Bob
\begin{eqnarray}\label{A9}
S^{A\rightarrow B}(T)&=&{\rm{max}}\bigg\{0,\frac{8}{\sqrt{3}}\big[|-q_{R}(e^{-\frac{\omega}{T}}+1)^{-\frac{1}{2}}\rho_{14}|^2-\mathcal{F}_a(T)-\mathcal{F}_b(T)\big],\nonumber\\
&&\frac{8}{\sqrt{3}}\big[|-q_{R}(e^{-\frac{\omega}{T}}+1)^{-\frac{1}{2}}\rho_{23}|^2-\mathcal{F}_c(T)-\mathcal{F}_b(T)\big]\bigg\},
\end{eqnarray}
and the quantum steering from Bob to Alice
\begin{eqnarray}\label{A10}
S^{B\rightarrow A}(T)&=&{\rm{max}}\bigg\{0,\frac{8}{\sqrt{3}}\big[|-q_{R}(e^{-\frac{\omega}{T}}+1)^{-\frac{1}{2}}\rho_{14}|^2-\mathcal{F}_a(T)+\mathcal{F}_b(T)\big],\nonumber\\
&&\frac{8}{\sqrt{3}}\big[|-q_{R}(e^{-\frac{\omega}{T}}+1)^{-\frac{1}{2}}\rho_{23}|^2-\mathcal{F}_c(T)+\mathcal{F}_b(T)\big]\bigg\},
\end{eqnarray}
where
\begin{eqnarray}\label{A11}
\mathcal{F}_{a}(T)&&=\frac{2-\sqrt{3}}{2}[(e^{-\frac{\omega}{T}}+1)^{-1}\rho_{11}+(1-|q_{R}|^{2})(e^{-\frac{\omega}{T}}+1)^{-1}\rho_{22}][(e^{\frac{\omega}{T}}+1)^{-1}\rho_{33}\nonumber\\
&&+\rho_{44}-(1-|q_{R}|^{2})(e^{-\frac{\omega}{T}}+1)^{-1}\rho_{44}]+\frac{2+\sqrt{3}}{2}[(e^{\frac{\omega}{T}}+1)^{-1}\rho_{11}+\rho_{22}\nonumber\\
&&-(1-|q_{R}|^{2})(e^{-\frac{\omega}{T}}+1)^{-1}\rho_{22}][(e^{-\frac{\omega}{T}}+1)^{-1}\rho_{33}+(1-|q_{R}|^{2})(e^{-\frac{\omega}{T}}+1)^{-1}\rho_{44}]\nonumber\\
&&+\frac{1}{4}[(e^{-\frac{\omega}{T}}+1)^{-1}\rho_{11}+(1-|q_{R}|^{2})(e^{-\frac{\omega}{T}}+1)^{-1}\rho_{22}+(e^{\frac{\omega}{T}}+1)^{-1}\rho_{33}+\rho_{44}\nonumber\\
&&-(1-|q_{R}|^{2})(e^{-\frac{\omega}{T}}+1)^{-1}\rho_{44}][(e^{\frac{\omega}{T}}+1)^{-1}\rho_{11}+\rho_{22}-(1-|q_{R}|^{2})(e^{-\frac{\omega}{T}}+1)^{-1}\rho_{22}\nonumber\\
&&+(e^{-\frac{\omega}{T}}+1)^{-1}\rho_{33}+(1-|q_{R}|^{2})(e^{-\frac{\omega}{T}}+1)^{-1}\rho_{44}],
\end{eqnarray}
\begin{eqnarray}\label{A12}
\mathcal{F}_{b}(T)&=&\frac{1}{4}[(e^{-\frac{\omega}{T}}+1)^{-1}\rho_{11}+(1-|q_{R}|^{2})(e^{-\frac{\omega}{T}}+1)^{-1}\rho_{22}-(e^{\frac{\omega}{T}}+1)^{-1}\rho_{33}-\rho_{44}\nonumber\\
&&+(1-|q_{R}|^{2})(e^{-\frac{\omega}{T}}+1)^{-1}\rho_{44}][(e^{\frac{\omega}{T}}+1)^{-1}\rho_{11}+\rho_{22}-(1-|q_{R}|^{2})(e^{-\frac{\omega}{T}}+1)^{-1}\rho_{22}\nonumber\\
&&-(e^{-\frac{\omega}{T}}+1)^{-1}\rho_{33}-(1-|q_{R}|^{2})(e^{-\frac{\omega}{T}}+1)^{-1}\rho_{44}],
\end{eqnarray}
\begin{eqnarray}\label{A13}
\mathcal{F}_{c}(T)&=&\frac{2+\sqrt{3}}{2}[(e^{-\frac{\omega}{T}}+1)^{-1}\rho_{11}+(1-|q_{R}|^{2})(e^{-\frac{\omega}{T}}+1)^{-1}\rho_{22}][(e^{\frac{\omega}{T}}+1)^{-1}\rho_{33}\nonumber\\
&&+\rho_{44}-(1-|q_{R}|^{2})(e^{-\frac{\omega}{T}}+1)^{-1}\rho_{44}]+\frac{2-\sqrt{3}}{2}[(e^{\frac{\omega}{T}}+1)^{-1}\rho_{11}+\rho_{22}\nonumber\\
&&-(1-|q_{R}|^{2})(e^{-\frac{\omega}{T}}+1)^{-1}\rho_{22}][(e^{-\frac{\omega}{T}}+1)^{-1}\rho_{33}+(1-|q_{R}|^{2})(e^{-\frac{\omega}{T}}+1)^{-1}\rho_{44}]\nonumber\\
&&+\frac{1}{4}[(e^{-\frac{\omega}{T}}+1)^{-1}\rho_{11}+(1-|q_{R}|^{2})(e^{-\frac{\omega}{T}}+1)^{-1}\rho_{22}+(e^{\frac{\omega}{T}}+1)^{-1}\rho_{33}+\rho_{44}\nonumber\\
&&-(1-|q_{R}|^{2})(e^{-\frac{\omega}{T}}+1)^{-1}\rho_{44}][(e^{\frac{\omega}{T}}+1)^{-1}\rho_{11}+\rho_{22}-(1-|q_{R}|^{2})(e^{-\frac{\omega}{T}}+1)^{-1}\rho_{22}\nonumber\\
&&+(e^{-\frac{\omega}{T}}+1)^{-1}\rho_{33}+(1-|q_{R}|^{2})(e^{-\frac{\omega}{T}}+1)^{-1}\rho_{44}].
\end{eqnarray}

\section{Ambiguity  of quantum teleportation and steering in Schwarzschild spacetime}
Ambiguity is related to the ordering criterion of the creation and annihilation operators and has gone unnoticed in fermionic quantum teleportation and steering in
Schwarzschild spacetime. Below, we briefly introduce ambiguity.
We consider a two-mode fermionic system associated with fermionic creation operators $a^{\dag}$ and $b^{\dag}$ acting on a vacuum state $|0\rangle$. Therefore, the relevant Hilbert space is four dimensional. The Hilbert space basis of our toy model can be represented as
\begin{eqnarray}\label{A14}
|00\rangle=|0\rangle,\quad |10\rangle=a^{\dag}|0\rangle,\quad |01\rangle=b^{\dag}|0\rangle,\quad |11\rangle=a^{\dag} b^{\dag}|0\rangle.
\end{eqnarray}
Based on this basis, we can endow the Hilbert space with a tensor product structure, which allows us to consider two qubits, where the first label corresponds to one qubit and the second label corresponds to the other qubit. We may change the quantum entanglement of state by making nonlocal changes to the basis: we swap the positions of $a^{\dag}$ and $b^{\dag}$ in Eq.(\ref{A14}) \cite{HL7,HL8,HL9,HL10,HL11}. The new basis can be obtained as
\begin{eqnarray}\label{A15}
|00\rangle'=|00\rangle,\quad |01\rangle'=|01\rangle,\quad |10\rangle'=|10\rangle,\quad |11\rangle'=b^{\dag}a^{\dag}|0\rangle=-|11\rangle.
\end{eqnarray}
Therefore, we can obtain a new basis in this specific case. Interestingly, choosing these two different types of bases may result in a separable state  being the Bell state.

In this paper, we use the operator ordering $\hat{a}^{\rm out}_{\bold k}\hat{b}^{\rm out}_{\bold {-k}}\hat{b}^{\rm in}_{\bold {-k}}\hat{a}^{\rm in}_{\bold k}$ to rewrite Eq.(\ref{S11}) as
\begin{eqnarray}\label{A16}
|0\rangle_{\rm U}'&=&\frac{1}{e^{-\frac{\omega}{T}}+1}|0000\rangle-\frac{1}{\sqrt{e^{\frac{\omega}{T}}+e^{-\frac{\omega}{T}}+2}}|0101\rangle\nonumber\\
&&+\frac{1}{\sqrt{e^{\frac{\omega}{T}}+e^{-\frac{\omega}{T}}+2}}|1010\rangle+\frac{1}{e^{\frac{\omega}{T}}+1}|1111\rangle.
\end{eqnarray}
However, Eq.(\ref{S13}) remains unchanged. Similarly, we can rewrite Eq.(\ref{S1}). Then,
we trace over its inaccessible modes and obtain
\begin{eqnarray}\label{Azq17}
\tilde{\rho}^{AB_{out}}= \left(\!\!\begin{array}{cccccccc}
\rho^{AB_{out}}_{11}& 0 & 0 & -\rho^{AB_{out}}_{14} &0 &\rho^{AB_{out}}_{16} &-\rho^{AB_{out}}_{17} &0\\
0& \rho^{AB_{out}}_{22} & 0 & 0 &\rho^{AB_{out}}_{25} &0 &0 &-\rho^{AB_{out}}_{28}\\
0& 0 & \rho^{AB_{out}}_{33} & 0 &-\rho^{AB_{out}}_{35} &0 &0 &-\rho^{AB_{out}}_{38}\\
-\rho^{AB_{out}}_{14}& 0 & 0 & \rho^{AB_{out}}_{44} &0 &-\rho^{AB_{out}}_{46} &-\rho^{AB_{out}}_{47} &0\\
0& \rho^{AB_{out}}_{25} & -\rho^{AB_{out}}_{35} & 0 &\rho^{AB_{out}}_{55} &0 &0 &-\rho^{AB_{out}}_{58}\\
\rho^{AB_{out}}_{16}& 0 & 0 & -\rho^{AB_{out}}_{46} &0 &\rho^{AB_{out}}_{66} &0 &0\\
-\rho^{AB_{out}}_{17}& 0 & 0 & -\rho^{AB_{out}}_{47} &0 &0 &\rho^{AB_{out}}_{77} &0\\
0& -\rho^{AB_{out}}_{28} & -\rho^{AB_{out}}_{38} & 0 &-\rho^{AB_{out}}_{58} &0 &0 &\rho^{AB_{out}}_{88}\\
\end{array}\!\!\right).
\end{eqnarray}
From  Eqs.(\ref{S2PO6}) and (\ref{Azq17}), we can see that the density matrix $\tilde{\rho}^{AB_{out}}$ is different from the density matrix $\rho^{AB_{out}}$. Then, we trace out the antifermionic mode of $\tilde{\rho}^{AB_{out}}$  outside the event horizon of the black hole and obtain
\begin{eqnarray}\label{Saq26}
\tilde{\rho}^{S}_{X}= \left(\!\!\begin{array}{cccccccc}
\rho^S_{11} & 0 & 0 & -\rho^S_{14} \\
0 & \rho^S_{22} & -\rho^S_{23} &0 \\
0 & -\rho^S_{23} & \rho^S_{33} & 0\\
-\rho^S_{14} & 0 & 0 & \rho^S_{44}
\end{array}\!\!\right).
\end{eqnarray}
From  Eqs.(\ref{S26}) and (\ref{Saq26}), we find that quantum state $\tilde{\rho}^{S}_{X}$ is the same as quantum state ${\rho}^{S}_{X}$.
Therefore, there is no ambiguous map in the quantum teleportation and steering for the same initial X-type state $\rho_{X}$  in the Schwarzschild black hole.

\begin{acknowledgments}
This work is supported by the National Natural
Science Foundation of China (Grant No. 12205133), LJKQZ20222315 and 2021BSL013.
\end{acknowledgments}

%\newpage


\begin{thebibliography}{99}
\bibitem{L1}
C. H. Bennett, G. Brassard, C. Cr\'{e}peau, R. Jozsa, A. Peres, and W. K. Wootters, Phys. Rev. Lett. {\bf70}, 1895 (1993).

\bibitem{L2}
D. Bouwmeester, J. W. Pan, K. Mattle, M. Eibl, H. Weinfurter, and A. Zeilinger, Nature {\bf390}, 575 (1997).

\bibitem{L3}
C. H. Bennett, G. Brassard, S. Popescu, B. Schumacher, J. A. Smolin, and W. K. Wootters, Phys. Rev. Lett. {\bf76}, 722 (1996).

\bibitem{H1}
N. Friis, A. R. Lee, K. Truong, C. Sab\'{\i}n, E. Solano, G. Johansson, and I. Fuentes, Phys. Rev. Lett. {\bf110}, 113602 (2013).

\bibitem{H2}
P. Lipka-Bartosik and P. Skrzypczyk, Phys. Rev. Lett. {\bf127}, 080502 (2021).

\bibitem{H3}
U. Marzolino and A. Buchleitner, Phys. Rev. A {\bf91}, 032316 (2015).

\bibitem{H4}
C. Noh, A. Chia, H. Nha, M. J. Collett, and H. J. Carmichael, Phys. Rev. Lett. {\bf102}, 230501 (2009).

\bibitem{H5}
X. Chen and K. W. C. Chan, Phys. Rev. A {\bf99}, 022334 (2019).

\bibitem{H6}
R. Fortes and G. Rigolin, Phys. Rev. A {\bf93}, 062330 (2016).

\bibitem{H7}
C. Q. Xu and D. L. Zhou, Phys. Rev. A {\bf106}, 012413 (2022).

\bibitem{H8}
X. M. Hu, C. Zhang, B. H. Liu, Y. Cai, X. J. Ye, Y. Guo, W. B. Xing, C. X. Huang, Y. F. Huang, C. F. Li, and G. C. Guo, Phys. Rev. Lett. {\bf125}, 230501 (2020).

\bibitem{H9}
S. Gangopadhyay, T. Wang, A. Mashatan, and S. Ghose, Phys. Rev. A {\bf106}, 052433 (2022).

\bibitem{H10}
B. Yoshida and N. Y. Yao, Phys. Rev. X {\bf9}, 011006 (2019).


\bibitem{L4}
H. M. Wiseman, S. J. Jones, and A. C. Doherty, Phys. Rev. Lett. {\bf98}, 140402 (2007).

\bibitem{L5}
S. J. Jones, H. M. Wiseman, and A. C. Doherty, Phys. Rev. A {\bf76}, 052116 (2007).

\bibitem{L6}
E. Schr\"{o}dinger, Proc. Cambridge Philos. Soc. {\bf31}, 555
(1935).

\bibitem{L7}
E. Schr\"{o}dinger, Proc. Cambridge Philos. Soc. {\bf32}, 446 (1936).

\bibitem{L8}
D. J. Saunders, S. J. Jones, H. M. Wiseman, and G. J. Pryde, Nat. Phys. {\bf6}, 845 (2010).

\bibitem{L9}
V. H\"{a}ndchen, T. Eberle, S. Steinlechner, A. Samblowski, T. Franz, R. F. Werner, and R. Schnabel, Nat. Photonics {\bf6}, 598 (2012).

\bibitem{L10}
S. Wollmann, N. Walk, A. J. Bennet, H. M. Wiseman, and G. J. Pryde, Phys. Rev. Lett. {\bf116}, 160403 (2016).

\bibitem{L11}
Y. Xiao, X. J. Ye, K. Sun, J. S. Xu, C. F. Li, and G. C. Guo, Phys. Rev. Lett. {\bf118}, 140404 (2017).

\bibitem{L12}
A. Ghosal, D. Das, S. Roy, and S. Bandyopadhyay, Phys. Rev. A {\bf101}, 012304 (2020).

\bibitem{L13}
B. P. Abbott $et$ $al.$, Phys. Rev. Lett. {\bf116}, 061102 (2016).

\bibitem{L14}
The Event Horizon Telescope Collaboration, Astrophys. J. Lett. {\bf875}, L1 (2019).

\bibitem{L15}
The Event Horizon Telescope Collaboration, Astrophys. J. Lett. {\bf875}, L2 (2019).

\bibitem{L16}
The Event Horizon Telescope Collaboration, Astrophys. J. Lett. {\bf875}, L3 (2019).

\bibitem{L17}
The Event Horizon Telescope Collaboration, Astrophys. J. Lett. {\bf875}, L4 (2019).

\bibitem{L18}
The Event Horizon Telescope Collaboration, Astrophys. J. Lett. {\bf875}, L5 (2019).

\bibitem{L19}
The Event Horizon Telescope Collaboration, Astrophys. J. Lett. {\bf875}, L6 (2019).

\bibitem{L20}
The Event Horizon Telescope Collaboration, Astrophys. J. Lett. {\bf930}, L12 (2022).

\bibitem{L21}
P. Xu, $et$ $al.$, Science {\bf366}, 132 (2019).

\bibitem{L22}
X. Busch, R. Parentani, Phys. Rev. D {\bf89}, 105024 (2014).

\bibitem{L23}
Z. Tian, J. Jing, A. Dragan, Phys. Rev. D {\bf95}, 125003 (2017).

\bibitem{L24}
Z. Tian, J. Du, Eur. Phys. J. C {\bf79}, 994 (2019).

\bibitem{L25}
J. Drori, Y. Rosenberg, D. Bermudez, Y. Silberberg, and U. Leonhardt, Phys. Rev. Lett. {\bf122}, 010404 (2019).

\bibitem{L26}
M. Isoard, N. Pavloff, Phys. Rev. Lett. {\bf124}, 060401 (2020).

\bibitem{L27}
Z. Tian, L. Wu, L. Zhang, J. Jing, J. Du, Phys. Rev. D {\bf106}, L061701 (2022).

\bibitem{L28}
J. Steinhauer, $et$ $al.$, Nature Communications {\bf13}, 2890 (2022).

\bibitem{L29}
C. Viermann, $et$ $al.$, Nature {\bf611}, 260 (2022).





\bibitem{L38}
S. Xu, X. K. Song, J. D. Shi, and L. Ye, Phys. Rev. D {\bf89}, 065022 (2014).

\bibitem{L39}
Q. Pan, J. Jing, Phys. Rev. D {\bf78}, 065015 (2008).

\bibitem{L40}
E. Mart\'{\i}n-Mart\'{\i}nez, L. J. Garay, and J. Le\'{o}n, Phys. Rev. D
{\bf82}, 064006 (2010).

\bibitem{L41}
J. Wang, Q. Pan, J. Jing, Phys. Lett. B {\bf692}, 202  (2010).

\bibitem{L42}
S. M. Wu, H. S. Zeng, Eur. Phys. J. C {\bf82}, 4 (2022).

\bibitem{L43}
J. Wang, H. Cao, J. Jing, and H. Fan, Phys. Rev. D {\bf93}, 125011 (2016).


\bibitem{L44}
S. Bhattacharya, N. Joshi, Phys. Rev. D {\bf105}, 065007 (2022).

\bibitem{L46}
I. Fuentes-Schuller, and R. B. Mann, Phys. Rev. Lett. {\bf 95},120404 (2005).

\bibitem{L47}
B. N. Esfahani, M. Shamirzaie, and M. Soltani, Phys. Rev. D {\bf84}, 025024 (2011).

\bibitem{L49}
S. M. Wu, Y. T. Cai, W. J. Peng, H. S. Zeng, Eur. Phys. J. C {\bf82}, 412 (2022).

\bibitem{L50}
P. M. Alsing, I. Fuentes-Schuller, R. B. Mann, and T. E. Tessier, Phys. Rev. A {\bf74}, 032326 (2006).

\bibitem{L51}
J. L. Huang, W. C. Gan, Y. Xiao, F. W. Shu, M. H. Yung, Eur. Phys. J. C {\bf78}, 545 (2018).

\bibitem{L52}
J. Wang, J. Jing, and H. Fan, Phys. Rev. D {\bf90}, 025032 (2014).

\bibitem{L53}
D. E. Bruschi, A. Datta, R. Ursin, T. C. Ralph, and I. Fuentes, Phys. Rev. D {\bf90}, 124001 (2014).

\bibitem{L54}
D. E. Bruschi, T. C. Ralph, I. Fuentes, T. Jennewein, and M. Razavi, Phys. Rev. D {\bf90}, 045041 (2014).

\bibitem{L55}
J. He, S. Xu, L. Ye, Phys. Lett. B {\bf756}, 278 (2016).

\bibitem{L56}
N. Hosseinidehaj, and R. Malaney, Phys. Rev. A {\bf91}, 022304 (2015).

\bibitem{L57}
F. Shahbazi, S. Haseli, H. Dolatkhaha and S. Salimi, JCAP {\bf10}, 47 (2020).

\bibitem{QLQ57}
S. M. Wu, H. S. Zeng, Eur. Phys. J. C {\bf82} 716 (2022).

\bibitem{L58}
S. Popescu, Phys. Rev. Lett {\bf72}, 797 (1994).

\bibitem{L59}
C. H. Bennett, D. P. DiVincenzo, J. A. Smolin, and W. K. Wootters, Phys. Rev. A {\bf54}, 3824 (1996).

\bibitem{L60}
P. Badziag, M. Horodecki, P. Horodecki, and R. Horodecki, Phys. Rev. A {\bf62}, 012311 (2000).

\bibitem{L62}
D. R. Brill and J. A. Wheeler, Rev. Mod. Phys. {\bf29}, 465 (1957).

\bibitem{L63}
J. Jing, Phys. Rev. D {\bf70}, 065004 (2004).

\bibitem{L64}
J. Wang, Q. Pan, J. Jing, Annals Phys. {\bf325}, 1190 (2010).

\bibitem{HS1}
T. Damoar, R. Ruffini, Phys. Rev. D {\bf14}, 332 (1976).

\bibitem{HL1}
D. E. Bruschi, J. Louko, E. Mart\'{\i}n-Mart\'{\i}nez, A. Dragan, and I. Fuentes, Phys. Rev. A {\bf82}, 042332 (2010).


\bibitem{HL2}
E. Mart\'{\i}n-Mart\'{\i}nez, D. Hosler, and M. Montero, Phys. Rev. A {\bf86}, 062307 (2012).

\bibitem{HL3}
J. Chang and Y. Kwon, Phys. Rev. A {\bf85}, 032302 (2012).

\bibitem{HL4}
E. Mart\'{\i}n-Mart\'{\i}nez and I. Fuentes, Phys. Rev. A {\bf83}, 052306 (2011).

\bibitem{HL5}
D. E. Bruschi, A. Dragan, I. Fuentes, and J. Louko, Phys. Rev. D {\bf86}, 025026 (2012).

\bibitem{HL6}
S. A. Hayward, R. Di Criscienzo, M. Nadalini, L. Vanzo, and S. Zerbini, Classical Quantum Gravity {\bf26}, 062001 (2009).

\bibitem{HL7}
M. Montero and E. Mart\'{\i}n-Mart\'{\i}nez, Phys. Rev. A {\bf83}, 062323 (2011).

\bibitem{HL8}
K. Br$\acute{a}$dler, and Roc\'{\i}o J$\acute{a}$uregui, Phys. Rev. A {\bf85}, 016301 (2012).

\bibitem{HL9}
M. Montero and E. Mart\'{\i}n-Mart\'{\i}nez, Phys. Rev. A {\bf85}, 016302 (2012).

\bibitem{HL10}
M. Montero and E. Mart\'{\i}n-Mart\'{\i}nez, Phys. Rev. A {\bf85}, 024301 (2012).

\bibitem{HL11}
N.Friis, A. R. Lee, D. E. Bruschi, Phys. Rev. A {\bf87}, 022338 (2013).





\bibitem{L66}
S. M. Hashemi Rafsanjani, M. Huber, C. J. Broadbent, and J. H. Eberly, Phys. Rev. A {\bf86}, 062303 (2012).

\bibitem{L67}
D. Das, S. Sasmal, and S. Roy, Phys. Rev. A {\bf99}, 052109 (2019).

\bibitem{L68}
K. Zhang, J. Wang, Phys. Rev. A {\bf104}, 042404 (2021).


\end{thebibliography}
\end{document}